\documentclass[11pt]{article}

\pdfoutput=1

\usepackage[T1]{fontenc}
\usepackage[latin9]{inputenc}
\usepackage[a4paper]{geometry}
\usepackage[active]{srcltx}
\usepackage{amsmath}
\usepackage{amssymb}
\usepackage{esint}

\makeatletter

\usepackage{textcomp}

\pdfoutput=1 

\usepackage{jheppub}

\usepackage{etoolbox}
    
    \patchcmd{\maketitle}{\@fpheader}{}{}{}

\usepackage{amsfonts}

\usepackage{tikz}

\usepackage{bbm}

\title{\boldmath Integrable systems with BMS$_{3}$ Poisson structure and the dynamics of locally flat spacetimes}

\author[a]{Oscar Fuentealba,}
\author[a,b]{Javier Matulich,}
\author[a]{Alfredo P\'erez,}
\author[c]{Miguel Pino,}
\author[a,d]{Pablo Rodr\'{i}guez,}
\author[a]{David Tempo,}
\author[a]{Ricardo Troncoso}

\affiliation[a]{Centro de Estudios Cient\'{i}ficos (CECs), Av. Arturo Prat 514, Valdivia, Chile}
\affiliation[b]{Universit\'e Libre de Bruxelles and International Solvay Institutes,
ULB-Campus Plaine CP231, 1050 Brussels, Belgium}
\affiliation[c]{Departamento de F\'{i}sica, Universidad de Santiago de Chile, Avenida Ecuador 3493, Estación Central, 9170124, Santiago, Chile}
\affiliation[d]{Departamento de F\'{i}sica, Universidad de Concepci\'on, Casilla 160-C, Concepci\'on, Chile}

\emailAdd{fuentealba@cecs.cl}
\emailAdd{jmatulic@ulb.ac.be}
\emailAdd{aperez@cecs.cl}
\emailAdd{miguel.pino.r@usach.cl}
\emailAdd{rodriguez@cecs.cl}
\emailAdd{tempo@cecs.cl}
\emailAdd{troncoso@cecs.cl}

\preprint{CECS-PHY-17/02}

\abstract{
We construct a hierarchy of integrable systems whose Poisson structure corresponds to the BMS$_{3}$ algebra, and then discuss its description in terms of the Riemannian geometry of locally flat spacetimes in three dimensions. 

The analysis is performed in terms of two-dimensional gauge fields for $isl(2,\mathbb{R})$, being isomorphic to the Poincar\'e algebra in 3D. Although the algebra is not semisimple, the formulation can still be carried out \textit{\`a la} Drinfeld-Sokolov because it admits a nondegenerate invariant bilinear metric. The hierarchy turns out to be bi-Hamiltonian, labeled by a nonnegative integer $k$, and defined through a suitable generalization of the Gelfand-Dikii polynomials. The symmetries of the hierarchy are explicitly found. For $k\geq 1$, the corresponding conserved charges span an infinite-dimensional Abelian algebra without central extensions, so that they are in involution; while in the case of $k=0$, they generate the BMS$_{3}$ algebra. In the special case of $k=1$, by virtue of a suitable field redefinition and time scaling, the field equations are shown to be equivalent to the ones of a specific type of the Hirota-Satsuma coupled KdV systems. For $k\geq 1$, the hierarchy also includes the so-called perturbed KdV equations as a particular case. A wide class of analytic solutions is also explicitly constructed for a generic value of $k$.

Remarkably, the dynamics can be fully geometrized so as to describe the evolution of spacelike surfaces embedded in locally flat spacetimes. Indeed, General Relativity in 3D can be endowed with a suitable set of boundary conditions, so that the Einstein equations precisely reduce to the ones of the hierarchy aforementioned. The symmetries of the integrable systems then arise as diffeomorphisms that preserve the asymptotic form of the spacetime metric, and therefore, they become Noetherian. The infinite set of conserved charges is then recovered from the corresponding surface integrals in the canonical approach.}

\makeatother

\begin{document}
\maketitle \flushbottom

\newpage{}

\section{Introduction}

The Galilean conformal algebra can be understood as the nonrelativistic
limit of the algebra of the conformal group (see, e.g. \cite{doi:10.1063/1.523670},
\cite{Bagchi:2009my}, \cite{Bagchi:2009pe}). In two spacetime dimensions,
the Galilean conformal algebra (GCA$_{2}$) is then obtained from
a suitable Inönü-Wigner contraction of two copies of the Virasoro
algebra, where the parameter of the contraction is the speed of light
($c\rightarrow\infty$). Remarkably, GCA$_{2}$ is isomorphic to the
Bondi-Metzner-Sachs algebra in three spacetime dimensions (BMS$_{3}$),
which spans the diffeomorphisms that preserve the asymptotic form
of the metric for General Relativity \cite{Ashtekar:1996cd}, \cite{Barnich:2006av},
\cite{Barnich:2010eb}, possibly endowed with parity-odd terms \cite{Barnich:2014cwa}.
The commutation relations are given by,
\begin{eqnarray}
i\left\{ \mathcal{J}_{m},\mathcal{J}_{n}\right\}  & = & \left(m-n\right)\mathcal{J}_{m+n}+2\pi c_{\mathcal{J}}\,m^{3}\delta_{m+n,0}\,,\nonumber \\
i\left\{ \mathcal{J}_{m},\mathcal{P}_{n}\right\}  & = & \left(m-n\right)\mathcal{P}_{m+n}+2\pi c_{\mathcal{P}}\,m^{3}\delta_{m+n,0}\,,\label{eq:BMSmodes-0}\\
i\left\{ \mathcal{P}_{m},\mathcal{P}_{n}\right\}  & = & 0\,,\nonumber 
\end{eqnarray}
with $m$ and $n$ arbitrary integers. The central extensions $c_{\mathcal{P}}$
and $c_{\mathcal{J}}$ are related to the Newton constant and to the
coupling of the parity-odd terms, respectively. The BMS$_{3}$ algebra
\eqref{eq:BMSmodes-0} is then described by the semi-direct sum of
a Virasoro algebra, spanned by $\mathcal{J}_{m}$, with the Abelian
ideal generated by $\mathcal{P}_{m}$. Note that the Poincaré algebra
in three dimensions is manifestly seen as the subalgebra of \eqref{eq:BMSmodes-0}
spanned by the subset of generators with $m$, $n=-1$,$0$,$1$ (after
suitable trivial shifts of $\mathcal{J}_{0}$ and $\mathcal{P}_{0}$).

The BMS$_{3}$ algebra also naturally arises in diverse contexts of
physical interest. For instance, it describes the worldsheet symmetries
of the bosonic sector in the tensionless limit of closed string theory
\cite{Bagchi:2013bga}, \cite{Bagchi:2015nca}, \cite{Mandal:2016lsa},
\cite{Casali:2016atr}, \cite{Bagchi:2016yyf}, \cite{Mandal:2016wrw},
\cite{Casali:2017zkz}, \cite{Bagchi:2017cte}, and it is then expected
to be relevant for the description of interacting higher spin fields
\cite{Francia:2002pt}, \cite{Francia:2006hp}, \cite{Sagnotti:2010at},
\cite{Polyakov:2009pk}, \cite{Polyakov:2010qs} (for a review, see
e.g. \cite{Bekaert:2010hw}). In two spacetime dimensions, the BMS$_{3}$
algebra also describes the symmetries of a ``flat analog'' of Liouville
theory \cite{Barnich:2012rz}, \cite{Barnich:2013yka}, while on Minkowski
spacetime in 3D, in the absence of central extensions, the algebra
\eqref{eq:BMSmodes-0} manifests itself through nonlocal symmetries
of a free massless Klein-Gordon field \cite{Batlle:2017llu}. The
algebra \eqref{eq:BMSmodes-0} also plays a key role in nonrelativistic
and flat holography \cite{Bagchi:2009my}, \cite{Barnich:2010eb},
\cite{Bagchi:2010eg}, \cite{Barnich:2012aw}, \cite{Gonzalez:2012nv},
\cite{Bagchi:2012yk}, \cite{Barnich:2012xq}, \cite{Bagchi:2012xr},
\cite{Barnich:2012rz}, \cite{Bagchi:2014iea}, \cite{Bagchi:2015wna},
\cite{Barnich:2015dvt}, \cite{Troessaert:2015syk}, \cite{Bagchi:2016geg},
\cite{Bagchi:2017cpu}. Furthermore, by virtue of a Sugawara-like
construction, their generators have been recently seen to emerge as
composite operators of the affine currents that describe the asymptotic
symmetries of the \textquotedblleft soft hairy\textquotedblright{}
boundary conditions in \cite{Afshar:2016wfy}, \cite{Grumiller:2016kcp},
\cite{Afshar:2016kjj}. In the sense of \cite{Hawking:2016msc}, \cite{Hawking:2016sgy},
\cite{Strominger:2017aeh}, this might shed light in the resolution
of the information loss paradox \cite{Hawking:2015qqa}. Similar results
also hold for the analysis of the near horizon symmetries of non-extremal
black holes, so that (twisted) warped conformal algebras also lead
to BMS$_{3}$ \cite{Donnay:2015abr}, \cite{Afshar:2015wjm}, \cite{Donnay:2016ejv}.
The minimal supersymmetric extension of the BMS$_{3}$ algebra has
been shown to generate the asymptotic symmetries of $\mathcal{N}=1$
supergravity in 3D \cite{Deser:1982sw}, \cite{Deser1984}, \cite{Marcus:1983hb},
for a suitable set of boundary conditions \cite{Barnich:2014cwa},
\cite{Barnich:2015sca}, and it is then isomorphic to the minimal
supersymmetric extension of GCA$_{2}$ \cite{Bagchi:2009ke}, \cite{Mandal:2010gx}
(see also \cite{Banerjee:2015kcx}). Supersymmetric extensions of
BMS$_{3}$ with $\mathcal{N}>1$ have also been discussed along diverse
lines in \cite{Casali:2016atr}, \cite{Bagchi:2016yyf}, \cite{Banerjee:2016nio},
\cite{Lodato:2016alv}, \cite{Mandal:2010gx}, \cite{Banerjee:2017gzj},
\cite{Basu:2017aqn}, \cite{Fuentealba:2017fck}. Interestingly, nonlinear
extensions of the BMS$_{3}$ algebra are known to exist when higher
spin bosonic or fermionic generators are included, see \cite{Afshar:2013vka},
\cite{Gonzalez:2013oaa}, \cite{Gary:2014ppa}, \cite{Matulich:2014hea},
\cite{Gonzalez:2014tba} and \cite{Fuentealba:2015jma}, \cite{Fuentealba:2015wza}
respectively. Further generalizations of the BMS$_{3}$ algebra can
also be found from suitable expansions of the Virasoro algebra \cite{Caroca:2017onr}.

One of the main purposes of our work is exploring whether the BMS$_{3}$
algebra could be further linked with some sort of integrable systems.
There are some hints that can be borrowed from CFTs in two dimensions
that suggest to look towards this direction. Indeed, it is known that
CFTs in 2D admit an infinite set of conserved charges that commute
between themselves, which can be constructed out from suitable nonlinear
combinations of the generators of the Virasoro algebra and their derivatives
(see e.g. \cite{DiFrancesco:639405}). Remarkably, these composite
operators turn out to be precisely the conserved charges of the KdV
equation, which also correspond to the Hamiltonians of the KdV hierarchy.
Therefore, since the BMS$_{3}$ algebra can be seen as a limiting
case of the conformal algebra in 2D, it is natural to wonder about
the possibility of performing a similar construction in that limit.
Specifically, one would like to know about the existence of an infinite
number of commuting conserved charges that could be suitably recovered
from the BMS$_{3}$ generators, as well as its possible relation with
some integrable system, or even to an entire hierarchy of them. Here
we show that this is certainly the case.

Furthermore, and noteworthy, the dynamics of this class of integrable
systems can be equivalently understood in terms of Riemannian geometry.
Indeed, following similar strategy as the one in \cite{Perez:2016vqo},
here we show that General Relativity without cosmological constant
in 3D can be endowed with a suitable set of boundary conditions, so
that in the reduced phase space, the Einstein equations precisely
agree with the ones of the hierarchy aforementioned. In other words,
the dynamics of our class of integrable system can be fully geometrized,
since it can be seen to emerge from the evolution of spacelike surfaces
embedded in locally flat spacetimes. As a remarkable consequence,
in the geometric picture, the symmetries of the integrable systems
correspond to diffeomorphisms that maintain the asymptotic form of
the spacetime metric, so that they manifestly become Noetherian. Hence,
the infinite set of conserved charges can be readily obtained from
the surface integrals associated to the asymptotic symmetries in the
canonical approach.

In the next section we explicitly construct dynamical (Hamiltonian)
systems whose Poisson structure corresponds to the BMS$_{3}$ algebra.
In order to analyze their symmetries and conserved charges, in section
\ref{sec:Drinfeld-Sokolov-formulation} we show how the Drinfeld-Sokolov
formulation can be adapted to our case, through the use of suitable
flat connections for $isl(2,\mathbb{R})$. Section \ref{sec:Hierarchy-of-integrable}
is devoted to a thorough construction and the analysis of an entire
bi-Hamiltonian hierarchy of integrable systems with BMS$_{3}$ Poisson
structure, labeled by a nonnegative integer $k$. We start with a
very simple dynamical system ($k=0$) from which the bi-Hamiltonian
structure can be naturally unveiled. The case of $k=1$ is described
in section \ref{subsec:Integrable-bi-Hamiltonian-systemk=00003D1},
where the Abelian infinite-dimensional symmetries and conserved charges
are explicitly identified in terms of a suitable generalization of
the Gelfand-Dikii polynomials (see also appendices \ref{sec:Lists}
and \ref{sec:Involution-of-the}). The equivalence between our field
equations and the ones of the Hirota-Satsuma coupled KdV system of
type ix, is shown to hold by virtue of a suitable field redefinition
and time scaling in section \ref{subsec:Equivalence-with-Hirota-Satsuma}.
The hierarchy of integrable systems with BMS$_{3}$ Poisson structure
is then explicitly discussed in section \ref{subsec:The-hierarchyk>1},
where it is shown that the so-called ``perturbed KdV equations''
are included as a particular case. Section \ref{subsec:Analytic-solutions}
is devoted to the construction of a wide interesting class of analytic
solutions for a generic value of the label of the hierarchy $k$.
In section \ref{sec:Geometrization-of-the}, we show how the dynamics
of the hierarchy of integrable systems can be fully geometrized in
terms of locally flat three-dimensional spacetimes. The deep link
with General Relativity in 3D is explicitly addressed. We conclude
with some remarks about possible extensions of our results in section
\ref{sec:Extensions}. Appendices \ref{sec:Lists}, \ref{sec:Involution-of-the}
and \ref{sec:Proof} are devoted to some technical remarks.

\section{Dynamical systems with BMS$_{3}$ Poisson structure\label{sec:Dynamical-systems-with-BMS3-Poisson-structure}}

In order to construct dynamical systems whose Poisson structure is
described by the BMS$_{3}$ algebra, let us consider two independent
dynamical fields, $\mathcal{J}=\mathcal{J}(t,\phi)$ and $\mathcal{\mathcal{P}=P}(t,\phi)$,
being defined on a cylinder whose coordinates range as $0\leq\phi<2\pi$,
and $-\infty<t<\infty$. The Poisson structure we look for can then
be defined in terms of the following operator
\begin{align}
\mathcal{D}^{(2)} & \equiv\left(\begin{array}{cc}
\mathcal{D}^{\left({\cal J}\right)} & \mathcal{D}^{\left({\cal P}\right)}\\
\mathcal{D}^{\left({\cal P}\right)} & 0
\end{array}\right)\;,\label{eq:D2}
\end{align}
where $\mathcal{D}^{\left(\mathcal{J}\right)}$ and $\mathcal{D}^{\left(\mathcal{P}\right)}$
stand for Schwarzian derivatives, given by 
\begin{eqnarray}
\mathcal{D}^{\left(\mathcal{J}\right)} & = & 2\mathcal{J}\partial_{\phi}+\partial_{\phi}\mathcal{J}-c_{\mathcal{J}}\partial_{\phi}^{3}\,,\nonumber \\
\mathcal{D}^{\left(\mathcal{P}\right)} & = & 2\mathcal{P}\partial_{\phi}+\partial_{\phi}\mathcal{P}-c_{\mathcal{P}}\partial_{\phi}^{3}\,,\label{eq:Schwarzian}
\end{eqnarray}
with arbitrary constants $c_{\mathcal{J}}$ and $c_{\mathcal{P}}$. 

The operator $\mathcal{D}^{(2)}$ in eq. \eqref{eq:D2} then allows
to write the Poisson brackets of two arbitrary functionals of the
dynamical fields, $F=F[\mathcal{J},\mathcal{P}]$ and $G=G[\mathcal{J},\mathcal{P}]$,
according to
\begin{equation}
\lbrace F,G\rbrace\equiv\int d\phi\left(\begin{array}{cc}
\frac{\delta F}{\delta\mathcal{J}} & \frac{\delta F}{\delta\mathcal{P}}\end{array}\right)\left(\begin{array}{cc}
\mathcal{D}^{\left({\cal J}\right)} & \mathcal{D}^{\left({\cal P}\right)}\\
\mathcal{D}^{\left({\cal P}\right)} & 0
\end{array}\right)\left(\begin{array}{c}
\frac{\delta G}{\delta\mathcal{J}}\\
\frac{\delta G}{\delta\mathcal{P}}
\end{array}\right)\,.\label{eq:brackets}
\end{equation}
 Therefore, the brackets of the dynamical fields read
\begin{eqnarray}
\lbrace\mathcal{J}\left(\phi\right),\mathcal{J}\left(\bar{\phi}\right)\rbrace & = & \mathcal{D}^{(\mathcal{J})}\delta\left(\phi-\bar{\phi}\right)\,,\nonumber \\
\lbrace\mathcal{J}\left(\phi\right),\mathcal{P}\left(\bar{\phi}\right)\rbrace & = & \mathcal{D}^{(\mathcal{P})}\delta\left(\phi-\bar{\phi}\right)\,,\label{eq:BMS}\\
\lbrace\mathcal{P}\left(\phi\right),\mathcal{P}\left(\bar{\phi}\right)\rbrace & = & 0\,,\nonumber 
\end{eqnarray}
so that expanding in Fourier modes as $X=\frac{1}{2\pi}\sum_{n}X_{n}e^{-in\phi}$,
the algebra of the Poisson brackets in \eqref{eq:BMS} precisely reduces
to the BMS$_{3}$ algebra in eq. \eqref{eq:BMSmodes-0}.

The field equations for the class of dynamical systems we were searching
for can then be readily defined as follows
\begin{equation}
\left(\begin{array}{c}
\dot{\mathcal{J}}\\
\dot{\mathcal{P}}
\end{array}\right)=\mathcal{D}^{\left(2\right)}\left(\begin{array}{c}
\mu_{\mathcal{J}}\\
\mu_{\mathcal{P}}
\end{array}\right)\,,\label{eq:Dynamical-system}
\end{equation}
where dot denotes the derivative in time, while $\mu_{\mathcal{J}}$
and $\mu_{\mathcal{P}}$ stand for arbitrary functions of the dynamical
fields and their derivatives along $\phi$. When these functions are
defined in terms of the variation of a functional $H=H[\mathcal{J},\mathcal{P}]$,
so that 
\begin{equation}
\mu_{\mathcal{J}}=\frac{\delta H}{\delta\mathcal{J}}\qquad,\qquad\mu_{\mathcal{P}}=\frac{\delta H}{\delta\mathcal{P}}\,,\label{eq:mus-2}
\end{equation}
the dynamical system is Hamiltonian; and hence, by virtue of \eqref{eq:brackets},
the field equations can be written as
\begin{equation}
\left(\begin{array}{c}
\dot{\mathcal{J}}\\
\dot{\mathcal{P}}
\end{array}\right)=\mathcal{D}^{\left(2\right)}\left(\begin{array}{c}
\mu_{\mathcal{J}}\\
\mu_{\mathcal{P}}
\end{array}\right)=\left(\begin{array}{c}
\left\{ \mathcal{J},H\right\} \\
\left\{ \mathcal{P},H\right\} 
\end{array}\right)\,.\label{eq:EQM}
\end{equation}
Note that unwrapping the angular coordinate to range as $-\infty<\phi<\infty$,
allows to extend this class of dynamical systems to $\mathbb{R}^{2}$,
provided that the fall-off of the dynamical fields $\mathcal{J}$,
$\mathcal{P}$ is sufficiently fast in order to get rid of boundary
terms. Hereafter, for the sake of simplicity, we will assume that
the dynamical systems are defined on a cylinder, with a single exception
for an interesting particular solution that is described in section
\ref{subsec:Particular-cases}.

\section{Zero-curvature formulation\label{sec:Drinfeld-Sokolov-formulation}}

In order to study the properties of the dynamical systems with BMS$_{3}$
Poisson structure that evolve according to eq. \eqref{eq:Dynamical-system},
including their symmetries and the corresponding conserved charges,
it turns out to be useful to reformulate them in terms of a flat connection
for a suitable Lie algebra (see e.g. \cite{das1989integrable,Dunajski:2010zz}).
In the standard approach of Drinfeld and Sokolov \cite{Drinfeld:1984qv},
the Lie algebra is assumed to be semisimple. Here we slightly extend
this approach in a sense explained right below. 

For our purposes, the relevant Lie algebra to be considered corresponds
to $isl(2,\mathbb{R})$, which is isomorphic to the Poincaré algebra
in 3D. Their commutation relations can then be written as
\begin{equation}
\left[J_{a},J_{b}\right]=\epsilon_{abc}J^{c}\quad,\quad\left[J_{a},P_{b}\right]=\epsilon_{abc}P^{c}\quad,\quad\left[P_{a},P_{b}\right]=0\,,\label{eq:isl(2R)}
\end{equation}
where $J_{a}$ stand for the generators of $sl(2,\mathbb{R})\simeq so(2,1)$.
It is worth emphasizing that the algebra \eqref{eq:isl(2R)} is not
semisimple; but nonetheless, it admits a non degenerate invariant
bilinear metric whose nonvanishing components read
\begin{equation}
\left\langle J_{a},J_{b}\right\rangle =c_{\mathcal{J}}\eta_{ab}\qquad,\qquad\left\langle J_{a},P_{b}\right\rangle =c_{\mathcal{P}}\eta_{ab}\,,\label{eq:bilinear metric}
\end{equation}
where $c_{\mathcal{J}}$ and $c_{\mathcal{P}}$ are arbitrary constants.\footnote{We choose the orientation according to $\epsilon_{012}=1$, while
the Minkowski metric $\eta_{ab}$ is assumed to be non-diagonal, whose
non-vanishing components are given by $\eta_{01}=\eta_{10}=\eta_{22}=1$. }

Hence, by virtue of \eqref{eq:bilinear metric}, the analysis of the
class of dynamical systems defined in section \ref{sec:Dynamical-systems-with-BMS3-Poisson-structure}
can still be performed \textit{\`a la} Drinfeld-Sokolov, provided
that the field equations are able to be reproduced in terms of a flat
connection for $isl(2,\mathbb{R})$. 

We then propose that the spacelike component of the $isl(2,\mathbb{R})$-valued
gauge field $a=a_{\mu}dx^{\mu}$ is given by 

\begin{equation}
a_{\phi}=J_{1}+\frac{1}{c_{\mathcal{P}}}\left[\left(\mathcal{J}-\frac{c_{\mathcal{J}}}{c_{\mathcal{P}}}\mathcal{P}\right)P_{0}+\mathcal{P}J_{0}\right]\,,\label{eq:a-phi}
\end{equation}
with $\mathcal{J}=\mathcal{J}(t,\phi)$ and $\mathcal{\mathcal{P}=P}(t,\phi)$,
while the timelike component reads

\begin{equation}
a_{t}=\Lambda\left(\mu_{\mathcal{J}},\mu_{\mathcal{P}}\right)\,,\label{eq:at}
\end{equation}
where 
\begin{align}
\Lambda\left(\mu_{\mathcal{J}},\mu_{\mathcal{P}}\right)= & \mu_{\mathcal{P}}P_{1}+\mu_{\mathcal{J}}J_{1}-\mu_{\mathcal{P}}\mbox{\ensuremath{'}}P_{2}-\mu_{\mathcal{J}}\mbox{\ensuremath{'}}J_{2}\nonumber \\
 & +\left(\frac{1}{c_{\mathcal{P}}}\mathcal{P}\mu_{\mathcal{J}}-\mu_{\mathcal{J}}\mbox{\ensuremath{''}}\right)J_{0}+\left[\frac{1}{c_{\mathcal{P}}}\left(\mathcal{J}-\frac{c_{\mathcal{J}}}{c_{\mathcal{P}}}\mathcal{P}\right)\mu_{\mathcal{J}}+\frac{1}{c_{\mathcal{P}}}\mathcal{P}\mu_{\mathcal{P}}-\mu_{\mathcal{P}}\mbox{\ensuremath{''}}\right]P_{0}\,.\label{eq:Lambda}
\end{align}
Here $\mu_{\mathcal{J}}$ and $\mu_{\mathcal{P}}$ can be assumed
to be given by some arbitrary functions of $\mathcal{J}$, $\mathcal{P}$,
and their derivatives along $\phi$ (being denoted by a prime here
and afterwards). Therefore, requiring the field strength for the gauge
field defined in \eqref{eq:a-phi} and \eqref{eq:at} to vanish, i.e.,

\begin{equation}
f=da+a^{2}=0\,,\label{eq:fchico}
\end{equation}
implies that the field equations for the dynamical system with BMS$_{3}$
Poisson structure in \eqref{eq:Dynamical-system} hold. It then goes
without saying that $a_{\phi}$ and $a_{t}$ correspond to a Lax pair.

\subsection{Symmetries and conserved charges\label{subsec:Symmetries-and-conserved}}

One of the advantages of formulating the field equations in terms
of a flat connection, is that the symmetries of the dynamical system
in \eqref{eq:Dynamical-system} turn out to correspond to gauge transformations,
$\delta_{\lambda}a=d\lambda+\left[a,\lambda\right]$, that preserve
the form of the gauge field defined through \eqref{eq:a-phi} and
\eqref{eq:at}.

Hence, requiring the form of the spacelike component of the connection
$a_{\phi}$ in \eqref{eq:a-phi} to be preserved under gauge transformations,
restricts the Lie-algebra-valued parameter to be of the form
\begin{equation}
\lambda=\Lambda\left(\varepsilon_{\mathcal{J}},\varepsilon_{\mathcal{P}}\right)\,,\label{eq:lambda-chico}
\end{equation}
where $\Lambda$ is precisely given by eq. \eqref{eq:Lambda}, but
now depends on two arbitrary functions $\varepsilon_{\mathcal{J}}=\varepsilon_{\mathcal{J}}(t,\phi)$
and $\varepsilon_{\mathcal{P}}=\varepsilon_{\mathcal{P}}(t,\phi)$,
while the dynamical fields have to transform according to
\begin{equation}
\left(\begin{array}{c}
\delta\mathcal{J}\\
\delta\mathcal{P}
\end{array}\right)=\mathcal{D}^{\left(2\right)}\left(\begin{array}{c}
\varepsilon_{\mathcal{J}}\\
\varepsilon_{\mathcal{P}}
\end{array}\right)\,.\label{eq:delta-P}
\end{equation}
Analogously, preserving the timelike component of the gauge field
$a_{t}$ in \eqref{eq:at}, implies that the transformation law of
the functions $\mu_{\mathcal{J}}$ and $\mu_{\mathcal{P}}$ is given
by
\begin{eqnarray}
\delta\mu_{\mathcal{J}} & = & \dot{\varepsilon}_{\mathcal{J}}+\varepsilon_{\mathcal{J}}\mu_{\mathcal{J}}\mbox{\ensuremath{\mbox{\ensuremath{'}}}}-\mu_{\mathcal{J}}\varepsilon{}_{\mathcal{J}}\mbox{\ensuremath{'}}\,,\label{eq:delta-mu-J}\\
\delta\mu_{\mathcal{P}} & = & \dot{\varepsilon}_{\mathcal{P}}+\varepsilon_{\mathcal{J}}\mu{}_{\mathcal{P}}\mbox{\ensuremath{'}}+\varepsilon_{\mathcal{P}}\mu{}_{\mathcal{J}}\mbox{\ensuremath{'}}-\mu_{\mathcal{J}}\varepsilon{}_{\mathcal{P}}\mbox{\ensuremath{'}}-\mu_{\mathcal{P}}\varepsilon{}_{\mathcal{J}}\mbox{\ensuremath{'}}\,.\label{eq:delta-mu-P}
\end{eqnarray}
 However, $\mu_{\mathcal{J}}$ and $\mu_{\mathcal{P}}$ generically
depend on the dynamical fields and their spatial derivatives, which
means that eqs. \eqref{eq:delta-mu-J}, \eqref{eq:delta-mu-P} actually
become a consistency condition to be fulfilled by the functions $\varepsilon_{\mathcal{J}}$
and $\varepsilon_{\mathcal{P}}$ that parametrize the symmetries of
the dynamical system.

In the case of Hamiltonian systems, $\mu_{\mathcal{J}}$ and $\mu_{\mathcal{P}}$
are determined by the corresponding functional variations of the Hamiltonian
as in \eqref{eq:mus-2}, and consequently, the consistency condition
for the functions $\varepsilon_{\mathcal{J}}$ and $\varepsilon_{\mathcal{P}}$,
that arises from \eqref{eq:delta-mu-J}, \eqref{eq:delta-mu-P}, can
be compactly written as
\begin{equation}
\left(\begin{array}{c}
\dot{\varepsilon}_{\mathcal{J}}\left(t,\phi\right)\\
\dot{\varepsilon}_{\mathcal{P}}\left(t,\phi\right)
\end{array}\right)=-\int d\varphi\left(\begin{array}{c}
\frac{\delta}{\delta\mathcal{J}\left(t,\phi\right)}\\
\frac{\delta}{\delta\mathcal{P}\left(t,\phi\right)}
\end{array}\right)\left(\mathcal{D}^{\left(2\right)}\left(\begin{array}{c}
\mu_{\mathcal{J}}\\
\mu_{\mathcal{P}}
\end{array}\right)\right)^{T}\left(\begin{array}{c}
\varepsilon_{\mathcal{J}}\\
\varepsilon_{\mathcal{P}}
\end{array}\right)\,.\label{eq:epsilon-punto}
\end{equation}
In sum, the functions that parametrize the symmetries of the Hamiltonian
system with BMS$_{3}$ Poisson structure must fulfill the consistency
condition in \eqref{eq:epsilon-punto}, which for an arbitrary choice
of Hamiltonian, implies that $\varepsilon_{\mathcal{J}}$ and $\varepsilon_{\mathcal{P}}$
generically acquire an explicit dependence on the dynamical fields
and their spatial derivatives.

\vspace{0.3cm}

The variation of the canonical generators associated to the symmetries
spanned by $\varepsilon_{\mathcal{J}}$, $\varepsilon_{\mathcal{P}}$
can then be readily found by virtue of eqs. \eqref{eq:bilinear metric},
\eqref{eq:a-phi} and \eqref{eq:lambda-chico}, which reduces to the
following simple expression 
\begin{equation}
\delta Q\left[\varepsilon_{\mathcal{J}},\varepsilon_{\mathcal{P}}\right]=-\int d\phi\left\langle \lambda\delta a_{\phi}\right\rangle =-\int d\phi\left(\varepsilon_{\mathcal{J}}\delta\mathcal{J}+\varepsilon_{\mathcal{P}}\delta\mathcal{P}\right)\,.\label{eq:canoncalQ}
\end{equation}
As a cross-check, it is simple to verify that the variation of the
canonical generators is conserved ($\delta\dot{Q}=0$) provided that
the consistency condition for the symmetry parameters in \eqref{eq:epsilon-punto}
is satisfied. 

Nevertheless, it must be highlighted that finding the explicit form
of the conserved charges $Q$ is not so simple, because it amounts
to know the general solution of the consistency condition for the
parameters in \eqref{eq:epsilon-punto}. Indeed, although the consistency
condition is linear in the parameters $\varepsilon_{\mathcal{J}}$,
$\varepsilon_{\mathcal{P}}$, the generic solution manifestly depends
on the dynamical fields and their spatial derivatives that fulfill
a nonlinear field equation. Thus, for a generic choice of the Hamiltonian,
solving eq. \eqref{eq:epsilon-punto} is actually an extremely difficult
task. 

However, for a generic Hamiltonian that is independent of the coordinates,
the conserved charges associated to translations along space and time
can be directly constructed. In fact, for a flat connection, diffeomorphisms
spanned by $\xi=\xi^{\mu}\partial_{\mu}$ are equivalent to gauge
transformations generated by $\lambda=-\xi^{\mu}a_{\mu}$, since $\mathcal{L}_{\xi}a=d\lambda+[a,\lambda]$.
Therefore, for $\xi=\partial_{\phi}$, the linear momentum on the
cylinder readily integrates as 

\begin{equation}
Q[\partial_{\phi}]=Q[-1,0]=\int d\phi\mathcal{J}\,.
\end{equation}

Analogously, the variation of the energy is obtained for $\xi=\partial_{t}$,
so that

\begin{equation}
\delta Q[\partial_{t}]=\delta Q[-\mu_{\mathcal{J}},-\mu_{\mathcal{P}}]=\int d\phi\left(\mu_{\mathcal{J}}\delta\mathcal{J}+\mu_{\mathcal{P}}\delta\mathcal{P}\right)\,,
\end{equation}
which by virtue of \eqref{eq:mus-2}, integrates as expected, i.e., 

\begin{equation}
Q[\partial_{t}]=H\,.\label{eq:Energy}
\end{equation}

Note that, generically, there might be additional nontrivial solutions
of eq. \eqref{eq:epsilon-punto} that would lead to further conserved
charges.

As a closing remark of this section, it must be emphasized that in
order to construct an integrable system with the BMS$_{3}$ Poisson
structure, one should at least specify the precise form of the Hamiltonian,
so that the general solution of the consistency condition for the
parameters in \eqref{eq:epsilon-punto} could be obtained. Explicit
examples of integrable systems of this sort that actually belong to
an infinite hierarchy of them are discussed in the next section.

\section{Hierarchy of integrable systems with BMS$_{3}$ Poisson structure\label{sec:Hierarchy-of-integrable}}

In this section we introduce a bi-Hamiltonian hierarchy of integrable
systems with BMS$_{3}$ Poisson structure in a constructive way. We
start from an extremely simple case, which nonetheless, possesses
the key ingredients in order to propose a precise nontrivial integrable
system of this type, that can be extended to an entire hierarchy labeled
by a nonnegative integer $k$. The contact with some known results
in the literature for certain particular cases is also addressed.
Furthermore, a wide class of analytic solutions are explicitly constructed
for an arbitrary representative of the hierarchy, including a couple
of simple and interesting particular examples. 

\subsection{Warming up with a simple dynamical system $\left(k=0\right)$\label{subsec:Warming-up-with}}

Let us begin with one of the simplest possible examples of a dynamical
system with BMS$_{3}$ Poisson structure. The field equations can
be obtained from \eqref{eq:EQM} with $\mu_{\mathcal{J}}=\mu_{\mathcal{J}}^{(0)}$
and $\mu_{\mathcal{P}}=\mu_{\mathcal{P}}^{(0)}$ constants, given
by
\begin{equation}
\left(\begin{array}{c}
\mu_{\mathcal{J}}^{\left(0\right)}\\
\mu_{\mathcal{P}}^{\left(0\right)}
\end{array}\right)=\left(\begin{array}{c}
1\\
a
\end{array}\right)\,,\label{eq:mu0}
\end{equation}
so that, according to \eqref{eq:mus-2}, the Hamiltonian is given
by $H=H^{(0)}$, with
\begin{equation}
H^{\left(0\right)}=\int d\phi\left(\mathcal{J}+a\mathcal{P}\right)\,.\label{eq:H(0)}
\end{equation}
The field equations then explicitly read 
\begin{eqnarray}
\dot{\mathcal{J}} & = & \mathcal{J}\text{\ensuremath{'}}+a\mathcal{P}\text{\ensuremath{'}}\,,\nonumber \\
\dot{\mathcal{P}} & = & \mathcal{P}\text{\ensuremath{'}}\,,\label{eq:EoMJP0}
\end{eqnarray}
which are trivially integrable. Indeed, the general solution of \eqref{eq:EoMJP0}
on the cylinder is described by left movers and it can be expressed
in terms of periodic functions $\mathcal{M=\mathcal{M}}(t+\phi)$
and $\mathcal{N=\mathcal{N}}(t+\phi)$, so that it reads 
\begin{eqnarray}
\mathcal{P} & = & \mathcal{M}\,,\nonumber \\
\mathcal{J} & = & \mathcal{N}+at\mathcal{M}'\,.\label{eq:sol-k=00003D0}
\end{eqnarray}
Besides, since the field equations are very simple in this case, the
consistency condition for the parameters of their symmetries in \eqref{eq:epsilon-punto}
becomes independent of the dynamical fields and their spatial derivatives,
which explicitly reduces to
\begin{eqnarray}
\dot{\varepsilon}_{\mathcal{J}} & = & \varepsilon{}_{\mathcal{J}}\mbox{\ensuremath{'}}\,,\nonumber \\
\dot{\varepsilon}_{\mathcal{P}} & = & \varepsilon{}_{\mathcal{P}}\mbox{\ensuremath{'}}+a\varepsilon{}_{\mathcal{J}}\mbox{\ensuremath{'}}\,.\label{eq:epsilonJP0}
\end{eqnarray}
Note that \eqref{eq:epsilonJP0} coincides with the field equations
in \eqref{eq:EoMJP0} for $\varepsilon_{\mathcal{J}}=\mathcal{P}$
and $\varepsilon_{\mathcal{P}}=\mathcal{J}$, and hence the general
solution of the consistency conditions for the parameters is also
given by chiral (left mover) functions as in \eqref{eq:sol-k=00003D0}.
Therefore, the variation of the canonical generators in \eqref{eq:canoncalQ}
readily integrates as
\begin{equation}
Q\left[\varepsilon_{\mathcal{J}},\varepsilon_{\mathcal{P}}\right]=-\int d\phi\left(\varepsilon_{\mathcal{J}}\mathcal{J}+\varepsilon_{\mathcal{P}}\mathcal{P}\right)\,.\label{eq:chargek=00003D0}
\end{equation}
The algebra of the conserved charges \eqref{eq:chargek=00003D0} can
then be directly obtained from their corresponding Poisson brackets.
As a shortcut, by virtue of
\begin{equation}
\left\{ Q\left[\varepsilon_{1}\right],Q\left[\varepsilon_{2}\right]\right\} =\delta_{\varepsilon_{2}}Q\left[\varepsilon_{1}\right]\,,\label{eq:QQdQ}
\end{equation}
the algebra can also be read from the transformation law of the fields
in \eqref{eq:delta-P}, and it is then found to be given precisely
by the BMS$_{3}$ algebra in \eqref{eq:BMS}, which once expanded
in modes reads as in eq. \eqref{eq:BMSmodes-0}.

It is worth highlighting that this simple dynamical system actually
turns out to be bi-Hamiltonian. This is so because the field equations
can be expressed in terms of two different Poisson structures, so
that \eqref{eq:EoMJP0} can be written as
\begin{equation}
\left(\begin{array}{c}
\dot{\mathcal{J}}\\
\dot{\mathcal{P}}
\end{array}\right)=\mathcal{D}^{\left(2\right)}\left(\begin{array}{c}
\mu_{\mathcal{J}}^{\left(0\right)}\\
\mu_{\mathcal{P}}^{\left(0\right)}
\end{array}\right)=\mathcal{D}{}^{(1)}\left(\begin{array}{c}
\mu_{\mathcal{J}}^{\left(1\right)}\\
\mu_{\mathcal{P}}^{\left(1\right)}
\end{array}\right)\,,\label{eq:EQM-1}
\end{equation}
where $\mathcal{D}^{(2)}$ is the BMS$_{3}$ one in \eqref{eq:D2},
while $\mathcal{D}^{(1)}$ stands for the ``canonical'' Poisson structure,
defined through the following differential operator
\begin{equation}
\mathcal{D}^{(1)}\equiv\left(\begin{array}{cc}
0 & \partial_{\phi}\\
\partial_{\phi} & 0
\end{array}\right)\,.\label{eq:D1}
\end{equation}
In \eqref{eq:EQM-1} the functions $\mu_{\mathcal{J}}^{(1)}$ and
$\mu_{\mathcal{P}}^{(1)}$ are then given by 
\begin{equation}
\left(\begin{array}{c}
\mu_{\mathcal{J}}^{\left(1\right)}\\
\mu_{\mathcal{P}}^{\left(1\right)}
\end{array}\right)=\left(\begin{array}{c}
\mathcal{P}\\
\mathcal{J}+a\mathcal{P}
\end{array}\right)\,,\label{eq:mus-1-1}
\end{equation}
and thus, according to \eqref{eq:mus-2}, the canonical Poisson structure
\eqref{eq:D1} is associated to the following Hamiltonian 
\begin{equation}
H^{\left(1\right)}=\int d\phi\left(\mathcal{J}\mathcal{P}+\frac{a}{2}\mathcal{P}^{2}\right)\,.\label{eq: H1}
\end{equation}

In sum, the analysis of this extremely simple dynamical system with
BMS$_{3}$ Poisson structure $\mathcal{D}^{(2)}$, being trivially
integrable, allows to unveil a naturally related Poisson structure
given by $\mathcal{D}^{(1)}$. The presence of both Poisson structures
turns out to be the key in order to proceed with the construction
of nontrivial integrable systems as well as an entire hierarchy associated
to them. Indeed, one can verify that any linear combination of $\mathcal{D}^{(1)}$
and $\mathcal{D}^{(2)}$ also defines a Poisson structure, which guarantees
the existence of the type of hierarchy that we are searching for. 

Furthermore, and remarkably, the simple dynamical system described
in this section can be seen to be equivalent to the Einstein equations
for the reduced phase space that is obtained from a suitable set of
boundary conditions for General Relativity in three spacetime dimensions,
including its extension with purely geometrical parity-odd terms in
the action. This is discussed in section \ref{sec:Geometrization-of-the}.
It is also worth pointing out that our field equations \eqref{eq:sol-k=00003D0},
in the case of $c_{\mathcal{J}}=a=0$, can be interpreted as the ones
of a compressible Euler fluid \cite{Penna:2017vms}.

In the next subsection we carry out the explicit construction and
the analysis of a simple, but nontrivial, integrable system with BMS$_{3}$
Poisson structure. 

\subsection{Integrable bi-Hamiltonian system $\left(k=1\right)$ \label{subsec:Integrable-bi-Hamiltonian-systemk=00003D1}}

The first nontrivial integrable system of our hierarchy is obtained
from \eqref{eq:EQM} with $\mu_{\mathcal{J}}=\mu_{\mathcal{J}}^{(1)}$
and $\mu_{\mathcal{P}}=\mu_{\mathcal{P}}^{(1)}$, where $\mu_{\mathcal{J}}^{(1)}$
and $\mu_{\mathcal{P}}^{(1)}$ are given by eq. \eqref{eq:mus-1-1},
so that the Hamiltonian corresponds to $H^{(1)}$ in \eqref{eq: H1}.
The field equations are then explicitly given by
\begin{eqnarray}
\dot{\mathcal{J}} & = & 3\mathcal{J}^{\prime}\mathcal{P}+3\mathcal{J}\mathcal{P}^{\prime}-c_{\mathcal{P}}\mathcal{J}^{\prime\prime\prime}-c_{\mathcal{J}}\mathcal{P}^{\prime\prime\prime}+a\left(3\mathcal{P}^{\prime}\mathcal{P}-c_{\mathcal{P}}\mathcal{P}^{\prime\prime\prime}\right)\,,\nonumber \\
\dot{\mathcal{P}} & = & 3\mathcal{P}^{\prime}\mathcal{P}-c_{\mathcal{P}}\mathcal{P}^{\prime\prime\prime}\,.\label{eq:EoMJP1}
\end{eqnarray}
Note that $\mathcal{P}$ evolves according to the KdV equation,\footnote{A common practice in the literature is rescaling the field and the
coordinates so that the KdV equation does not depend on $c_{\mathcal{P}}$
(see, e.g., \cite{das1989integrable,olver2000applications}). However,
as explained in section \ref{subsec:Symmetries}, for our purposes,
and for the sake of simplicity, keeping $c_{\mathcal{P}}$ explicitly
in the field equations turns out to be very useful and convenient.} while the remaining equation is linear in $\mathcal{J}$, with an
inhomogeneous source term that is entirely determined by $\mathcal{P}$
and their spatial derivatives.

The field equations in \eqref{eq:EoMJP1} can also be seen to arise
from a bi-Hamiltonian system with the same BMS$_{3}$ and canonical
Poisson structures given by \eqref{eq:D2} and \eqref{eq:D1}, respectively.
Indeed, they can be written as
\begin{equation}
\left(\begin{array}{c}
\dot{\mathcal{J}}\\
\dot{\mathcal{P}}
\end{array}\right)=\mathcal{D}{}^{(2)}\left(\begin{array}{c}
\mu_{\mathcal{J}}^{\left(1\right)}\\
\mu_{\mathcal{P}}^{\left(1\right)}
\end{array}\right)=\mathcal{D}{}^{(1)}\left(\begin{array}{c}
\mu_{\mathcal{J}}^{\left(2\right)}\\
\mu_{\mathcal{P}}^{\left(2\right)}
\end{array}\right)\,,\label{eq:EQM-2-1}
\end{equation}
where the functions $\mu_{\mathcal{J}}^{\left(2\right)}$ and $\mu_{\mathcal{P}}^{\left(2\right)}$
are given by
\begin{equation}
\left(\begin{array}{c}
\mu_{\mathcal{J}}^{\left(2\right)}\\
\mu_{\mathcal{P}}^{\left(2\right)}
\end{array}\right)=\left(\begin{array}{c}
\frac{3}{2}\mathcal{P}^{2}-c_{\mathcal{P}}\mathcal{P}^{\prime\prime}\\
3\mathcal{J}\mathcal{P}-c_{\mathcal{P}}\mathcal{J}^{\prime\prime}-c_{\mathcal{J}}\mathcal{P}^{\prime\prime}+a\left(\frac{3}{2}\mathcal{P}^{2}-c_{\mathcal{P}}\mathcal{P}^{\prime\prime}\right)
\end{array}\right)\,,
\end{equation}
which can be obtained from the functional derivatives of a different
Hamiltonian, as in \eqref{eq:mus-2}, that reads
\begin{equation}
H^{\left(2\right)}=\int d\phi\left[\frac{3}{2}\mathcal{P}^{2}\mathcal{J}-c_{\mathcal{P}}\mathcal{P}^{\prime\prime}\mathcal{J}+\frac{c_{\mathcal{J}}}{2}\mathcal{P}^{\prime2}+a\left(\frac{1}{2}\mathcal{P}^{3}+\frac{c_{\mathcal{P}}}{2}\mathcal{P}^{\prime2}\right)\right]\,,\label{eq:H2}
\end{equation}
being clearly conserved.

\subsubsection{Symmetries\label{subsec:Symmetries}}

As explained in section \ref{subsec:Symmetries-and-conserved}, in
order to find the remaining conserved quantities, it is necessary
to find the general solution of the consistency conditions in \eqref{eq:epsilon-punto}
for the functions $\varepsilon_{\mathcal{J}}$ and $\varepsilon_{\mathcal{P}}$
that parametrize the symmetries of the field equations. In this case
($k=1$), the consistency conditions in \eqref{eq:epsilon-punto},
with $\mu_{\mathcal{J}}$ and $\mu_{\mathcal{P}}$ given by eq. \eqref{eq:mus-1-1},
explicitly reduce to
\begin{eqnarray}
\dot{\varepsilon}_{\mathcal{J}} & = & 3\mathcal{P}\varepsilon_{{\cal J}}^{\prime}-c_{\mathcal{P}}\varepsilon_{{\cal J}}^{\prime\prime\prime}\,,\nonumber \\
\dot{\varepsilon}_{\mathcal{P}} & = & 3\mathcal{J}\varepsilon_{{\cal J}}^{\prime}+3\mathcal{P}\varepsilon_{{\cal P}}^{\prime}-c_{\mathcal{P}}\varepsilon_{{\cal P}}^{\prime\prime\prime}-c_{\mathcal{J}}\varepsilon_{{\cal J}}^{\prime\prime\prime}+a\left(3\mathcal{P}\varepsilon_{{\cal J}}^{\prime}-c_{\mathcal{P}}\varepsilon_{{\cal J}}^{\prime\prime\prime}\right)\,.\label{eq:epsilonJP1}
\end{eqnarray}
Note that the equations in \eqref{eq:epsilonJP1} are linear for the
parameters $\varepsilon_{\mathcal{J}}$ and $\varepsilon_{\mathcal{P}}$.
However, finding their solution is not that simple because their coefficients
depend on $\mathcal{J}$ and $\mathcal{P}$, who evolve according
to the nonlinear field equations in \eqref{eq:EoMJP1}. Nevertheless,
if one assumes that the parameters are given by local functions of
the dynamical fields and their spatial derivatives, the general solution
of the consistency conditions \eqref{eq:epsilonJP1} is found to be
given by a linear combination of two independent arrays, $K^{(j)}$
and $\tilde{K}^{(j)}$, that stand for a suitable generalization of
the Gelfand-Dikii polynomials. The solution can then be written as
\begin{equation}
\left(\begin{array}{c}
\varepsilon_{\mathcal{J}}\\
\varepsilon_{\mathcal{P}}
\end{array}\right)=\sum_{j=0}^{\infty}\left[\eta_{j}K^{(j)}+\tilde{\eta}_{j}\tilde{K}^{(j)}\right]\,,\label{eq:sum-polynomials}
\end{equation}
where $\eta_{j}$ and $\tilde{\eta}_{j}$ are arbitrary constants,
and both generalized polynomials $K^{(j)}$ and $\tilde{K}^{(j)}$
fulfill the same recursive relationship, given by

\begin{equation}
\mathcal{D}^{(1)}K{}^{(i+1)}=\mathcal{D}^{(2)}K^{(i)}\,.\label{eq:recrelation}
\end{equation}
If the initial seeds of the independent arrays are chosen as
\begin{equation}
K^{(0)}=\left(\begin{array}{c}
0\\
1
\end{array}\right)\quad,\quad\tilde{K}^{(0)}=\left(\begin{array}{c}
1\\
0
\end{array}\right)\,,
\end{equation}
the recursion relation \eqref{eq:recrelation} then implies that the
remaining ones are given by

\begin{equation}
K^{(n)}=\left(\begin{array}{c}
0\\
R^{\left(n\right)}
\end{array}\right)\quad,\quad\tilde{K}^{(n)}=\left(\begin{array}{c}
R^{\left(n\right)}\\
T^{\left(n\right)}
\end{array}\right)\,,
\end{equation}
where $R^{\left(n\right)}$ stand for the standard Gelfand-Dikii polynomials,
while $T^{\left(n\right)}$ correspond to a different set of polynomials
that fulfill the following recursion relationships 
\begin{eqnarray}
\partial_{\phi}R^{\left(n+1\right)} & = & \mathcal{D}^{\left(\mathcal{P}\right)}R^{\left(n\right)}\,,\label{eq:Gelfan-1}\\
\partial_{\phi}T^{\left(n+1\right)} & = & \mathcal{D}^{\left(\mathcal{P}\right)}T^{\left(n\right)}+\mathcal{D}^{\left(\mathcal{J}\right)}R^{\left(n\right)}\,.\label{Gelfan_Tempo-1}
\end{eqnarray}
Remarkably, both sets of polynomials can be obtained from the variation
of two independent functionals, $H_{\text{KdV}}^{\left(n\right)}\left[\mathcal{P}\right]$
and $\tilde{H}^{\left(n\right)}\left[\mathcal{P},\mathcal{J}\right]$,
so that 
\begin{eqnarray}
R^{\left(n\right)} & = & \frac{\delta H_{\text{KdV}}^{\left(n\right)}\left[\mathcal{P}\right]}{\delta\mathcal{P}}=\frac{\delta\tilde{H}^{\left(n\right)}\left[\mathcal{P},\mathcal{J}\right]}{\delta\mathcal{J}}\,,\label{eq:Rn-1}\\
T^{\left(n\right)} & = & \frac{\delta\tilde{H}^{\left(n\right)}\left[\mathcal{P},\mathcal{J}\right]}{\delta\mathcal{P}}\,,\label{eq:Tn-1}
\end{eqnarray}
where $H_{\text{KdV}}^{\left(n\right)}$ stands for $n$-th conserved
quantity of the KdV equation, while $\tilde{H}^{\left(n\right)}\left[\mathcal{P},\mathcal{J}\right]$
depends linearly on $\mathcal{J}$ and it is given by
\begin{equation}
\tilde{H}^{\left(n\right)}\left[\mathcal{P},\mathcal{J}\right]=c_{\mathcal{J}}\frac{\partial H_{\text{KdV}}^{\left(n\right)}\left[\mathcal{P}\right]}{\partial c_{\mathcal{P}}}+\int d\phi{\cal J}\frac{\delta H_{\text{KdV}}^{\left(n\right)}\left[\mathcal{P}\right]}{\delta\mathcal{P}}\,.\label{eq:Htilde}
\end{equation}
Therefore, the generalized polynomials can also be compactly defined
as
\begin{equation}
K^{(n)}=\left(\begin{array}{c}
\frac{\delta}{\delta\mathcal{J}}\\
\frac{\delta}{\delta\mathcal{P}}
\end{array}\right)H_{\text{KdV}}^{(n)}\quad,\quad\tilde{K}^{(n)}=\left(\begin{array}{c}
\frac{\delta}{\delta\mathcal{J}}\\
\frac{\delta}{\delta\mathcal{P}}
\end{array}\right)\tilde{H}^{(n)}\,.\label{eq:Kas(n)}
\end{equation}

An explicit list of the first six polynomials $R^{(n)}$ and $T^{(n)}$,
with their corresponding functionals $H_{\text{KdV}}^{\left(n\right)}$
and $\tilde{H}^{\left(n\right)}$ is given in appendix \ref{sec:Lists}.

\subsubsection{Conserved charges\label{subsec:Conserved-charges}}

Since the general form of the parameters that describe the symmetries
of the field equations has been explicitly found to be given by \eqref{eq:sum-polynomials},
by virtue of the fact that the generalized polynomials come from the
functional derivatives of suitable functionals as in \eqref{eq:Kas(n)},
the variation of the canonical generators in \eqref{eq:canoncalQ}
reduces to
\begin{equation}
\delta Q\left[\eta,\tilde{\eta}\right]=-\sum_{j=0}^{\infty}\int d\phi\left[\eta_{j}\frac{\delta H_{\text{KdV}}^{\left(j\right)}}{\delta\mathcal{P}}\delta\mathcal{P}+\tilde{\eta}_{j}\left(\frac{\delta\tilde{H}^{\left(j\right)}}{\delta\mathcal{P}}\delta\mathcal{P}+\frac{\delta\tilde{H}^{\left(j\right)}}{\delta\mathcal{J}}\delta\mathcal{J}\right)\right]\,,
\end{equation}
which then readily integrates as
\begin{equation}
Q\left[\eta,\tilde{\eta}\right]=-\sum_{j=0}^{\infty}\left(\eta_{j}H_{\text{KdV}}^{\left(j\right)}+\tilde{\eta}_{j}\tilde{H}^{\left(j\right)}\right)\;.\label{eq:Q-1}
\end{equation}

Therefore, we have explicitly found two infinite independent towers
of conserved quantities, being spanned by $H_{\text{KdV}}^{\left(j\right)}$
and $\tilde{H}^{\left(j\right)}$, which by virtue of the recursion
relation in \eqref{eq:recrelation}, turn out to be in involution
for both Poisson structures $\mathcal{D}^{(2)}$ and $\mathcal{D}^{(1)}$,
i.e., 
\begin{equation}
\left\{ Q[\eta,\tilde{\eta}],Q[\zeta,\tilde{\zeta}]\right\} _{(2)}=\left\{ Q[\eta,\tilde{\eta}],Q[\zeta,\tilde{\zeta}]\right\} _{(1)}=0\,,\label{eq:involution}
\end{equation}
where the subscripts for the brackets in \eqref{eq:involution} stand
for the corresponding Poisson structures.\footnote{For an explicit proof of the involution of the conserved charges see
appendix \ref{sec:Involution-of-the}.}

Note that the pair of Hamiltonians that yield the same field equations
in \eqref{eq:EQM-2-1}, given by \eqref{eq: H1} and \eqref{eq:H2},
can then be written as
\begin{equation}
H^{(1)}=\tilde{H}^{(1)}+aH_{\text{KdV}}^{(1)}\quad,\quad H^{(2)}=\tilde{H}^{(2)}+aH_{\text{KdV}}^{(2)}\,.\label{eq:H1andH2}
\end{equation}

In sum, as pointed out in the introduction, eq. \eqref{eq:Q-1} turns
out to be an explicit realization of the infinite set of commuting
conserved charges that is constructed out from precise nonlinear combinations
of the generators of the BMS$_{3}$ algebra and their spatial derivatives.
As discussed in section \ref{subsec:The-hierarchyk>1}, this provides
the basis to extend this integrable system to an entire hierarchy.

\subsubsection{Remarks on some additional symmetries \label{RemarksSymm}}

Apart from the infinite set of symmetries spanned by \eqref{eq:delta-P},
with $\varepsilon_{\mathcal{J}}$ and $\varepsilon_{\mathcal{P}}$
given by \eqref{eq:sum-polynomials}, the field equations \eqref{eq:EoMJP1}
can also be seen to be invariant under Galilean and anisotropic scale
transformations:

\emph{Galilean transformations.-} They are parametrized by a single
constant velocity parameter $v_{0}$, so that the coordinates and
the fields transform according to
\begin{equation}
\bar{\phi}=\phi+v_{0}t\qquad,\qquad\bar{t}=t\,,\label{eq:Galilean-boost}
\end{equation}
and
\begin{equation}
\bar{\mathcal{P}}=\mathcal{P}-\frac{v_{0}}{3}\qquad,\qquad\bar{\mathcal{J}}=\mathcal{J}+\frac{a}{3}v_{0}\,,
\end{equation}
respectively. 

\emph{Anisotropic scaling of Lifshitz type.-} This symmetry is defined
through a constant parameter $\sigma$, and it is generated by 
\begin{equation}
\bar{t}=\sigma^{3}t\quad,\quad\bar{\phi}=\sigma\phi\quad,\quad\left(\begin{array}{c}
\bar{\mathcal{J}}\\
\bar{\mathcal{P}}
\end{array}\right)=\sigma^{-2}\left(\begin{array}{c}
\mathcal{J}\\
\mathcal{P}
\end{array}\right)\,,\label{eq:Scaling-transf}
\end{equation}
which corresponds to an anisotropic scaling of Lifshitz type with
dynamical exponent $z=3$ (for a deeper discussion on anisotropic
scaling of Lifshitz type, see e.g. \cite{Taylor:2008tg,Bertoldi:2009vn,Bertoldi:2009dt,Hartnoll:2009sz,DHoker:2010zpp,Hartnoll:2011fn,Gonzalez:2011nz,Hartnoll:2015faa,Taylor:2015glc,Perez:2016vqo}).
For later purposes, it is worth noting that both sets of conserved
quantities, $H_{\text{KdV}}^{(n)}$ and $\tilde{H}^{(n)}$, scale
under \eqref{eq:Scaling-transf} according to
\begin{equation}
\bar{H}^{\left(n\right)}=\sigma^{-(2n+1)}H^{\left(n\right)}\,.\label{eq:Hnscaling}
\end{equation}

\subsubsection{Equivalence with the Hirota-Satsuma coupled KdV system of type ix\label{subsec:Equivalence-with-Hirota-Satsuma}}

Here we show that the field equations of the bi-Hamiltonian integrable
system with BMS$_{3}$ and canonical Poisson structures, described
by \eqref{eq:EoMJP1}, can be seen to be equivalent to a particular
class of a generalization of the Hirota-Satsuma coupled KdV system
\cite{Hirota:1981wb}. Specifically, according to the classification
in \cite{Sakovich}, the equivalence is shown to hold for the field
equations of type ix, which have been shown to be integrable through
a method that differs from the one we have used above. The equivalence
can be seen as follows.

If one changes the dynamical fields and rescale time according to
\begin{equation}
u=\frac{1}{4c_{\mathcal{P}}}\mathcal{P}\quad,\quad v=\frac{1}{4}\left(\mathcal{J}+a\mathcal{P}\right)\quad,\quad\tau=-c_{\mathcal{P}}t\,,\label{eq:HSvariables}
\end{equation}
the field equations in \eqref{eq:EoMJP1} read
\begin{eqnarray}
\partial_{\tau}v & = & -12uv\text{\ensuremath{'}}-12vu\text{\ensuremath{'}}+v\text{\ensuremath{'''}}+\gamma u\text{\ensuremath{'''}}\,,\label{eq:vtau}\\
\partial_{\tau}u & = & -12uu\text{\ensuremath{'}}+u\text{\ensuremath{'''}}\,,\label{eq:utau-1}
\end{eqnarray}
with
\begin{equation}
\gamma\equiv c_{\mathcal{J}}+ac_{\mathcal{P}}\,.\label{eq:lambda}
\end{equation}
The equations in \eqref{eq:vtau}, \eqref{eq:utau-1} then turn out
to be precisely the ones of type ix in \cite{Sakovich}. Thus, at
the level of the field equations there are actually only two inequivalent
cases. The generic one corresponds to $\gamma=1$, since the field
equation in \eqref{eq:vtau} can always be brought to this form provided
that $v$ is rescaled as $v\rightarrow\gamma v$. The remaining case
is described by $\gamma=0$, which is also known in the literature
as ``perturbed KdV'' (see e.g. \cite{0305-4470-16-16-015}, \cite{Ma:1996nv},
\cite{MA199649}, \cite{KalkanliKarasu:1997jm}, \cite{Ma:1999nv},
\cite{1742-6596-490-1-012024}, \cite{Restuccia:2016gts}).

Note that for $\gamma=0$ ($c_{\mathcal{J}}=-ac_{\mathcal{P}}$),
configurations with $v=0$ (${\cal J}=-a{\cal P}$) are devoid of
energy, since $H=H^{\left(0\right)}$ in \eqref{eq:H(0)} vanishes.
Nonetheless, they are generically endowed with both towers of conserved
charges $H_{\text{KdV}}^{(j)}$ and $\tilde{H}^{(j)}$.

\medskip{}

In sum, we would like to emphasize that both cases $\left(\gamma=0,1\right)$,
have now been shown to be endowed with radically inequivalent bi-Hamiltonian
structures depending on whether the BMS$_{3}$ central charge $c_{\mathcal{J}}$
vanishes or not. Therefore, these structures provide all what is needed
in order to generalize the integrable system discussed in this subsection
to a hierarchy of them that is labeled by a nonnegative integer $k$.

\subsection{The hierarchy $\left(k\geq0\right)$\label{subsec:The-hierarchyk>1}}

The results obtained in section \ref{subsec:Integrable-bi-Hamiltonian-systemk=00003D1}
ensure that a hierarchy of bi-Hamiltonian integrable systems with\textbf{
}BMS$_{3}$ and canonical Poisson structures, given by $\mathcal{D}^{(2)}$
and $\mathcal{D}^{(1)}$, can be readily constructed out from choosing
any of their Hamiltonians to be given by an arbitrary linear combination
of the independent conserved charges $H_{\text{KdV}}^{(n)}$ and $\tilde{H}^{(n)}$
in \eqref{eq:Htilde}. Hereafter we focus in the subclass of them
that possesses well-defined scaling properties. In order to achieve
this task, we recall that if one rescales the spacelike coordinate
and the fields as in \eqref{eq:Scaling-transf}, then both sets of
conserved charges scale according to \eqref{eq:Hnscaling}. Therefore,
the most general combination that possesses the suitable scaling properties
that we look for is described by a Hamiltonian of the form
\begin{equation}
H^{(k)}=\tilde{H}^{(k)}+aH_{\text{KdV}}^{(k)}\,,\label{eq:Hk}
\end{equation}
with $a$ constant and any fixed value of the integer $k\geq0$. 

The field equations of the corresponding hierarchy of integrable systems
then read

\begin{equation}
\left(\begin{array}{c}
\dot{\mathcal{J}}\\
\dot{\mathcal{P}}
\end{array}\right)=\mathcal{D}^{\left(2\right)}\left(\begin{array}{c}
\mu_{\mathcal{J}}^{\left(k\right)}\\
\mu_{\mathcal{P}}^{\left(k\right)}
\end{array}\right)\,,\label{eq:Feqsk}
\end{equation}
 with
\begin{equation}
\mu_{\mathcal{J}}^{(k)}=\frac{\delta H^{(k)}}{\delta\mathcal{J}}\quad\mbox{and}\quad\mu_{\mathcal{P}}^{(k)}=\frac{\delta H^{(k)}}{\delta\mathcal{P}}\,.\label{eq:musk}
\end{equation}

The hierarchy defined through \eqref{eq:Feqsk} is clearly bi-Hamiltonian,
since by virtue of the recursion relationship in \eqref{eq:recrelation},
the field equations can also be expressed as
\begin{equation}
\left(\begin{array}{c}
\dot{\mathcal{J}}\\
\dot{\mathcal{P}}
\end{array}\right)=\mathcal{D}^{\left(2\right)}\left(\begin{array}{c}
\mu_{\mathcal{J}}^{\left(k\right)}\\
\mu_{\mathcal{P}}^{\left(k\right)}
\end{array}\right)=\mathcal{D}^{(1)}\left(\begin{array}{c}
\mu_{\mathcal{J}}^{\left(k+1\right)}\\
\mu_{\mathcal{P}}^{\left(k+1\right)}
\end{array}\right)\,.\label{eq:hierarchy_eq}
\end{equation}
Furthermore, by construction, the field equations for any representative
of the hierarchy turn out to be invariant under anisotropic scaling
transformations given by
\begin{equation}
t\rightarrow\sigma^{z}t\quad,\quad\phi\rightarrow\sigma\phi\quad,\quad\left(\begin{array}{c}
\mathcal{J}\\
\mathcal{P}
\end{array}\right)\rightarrow\sigma^{-2}\left(\begin{array}{c}
\mathcal{J}\\
\mathcal{P}
\end{array}\right)\,,\label{eq:Lifshitz-z-scaling}
\end{equation}
which is of Lifshitz type, and characterized by a dynamical exponent
$z=2k+1$.

The field equations of the hierarchy can also be explicitly written
in terms of the polynomials $T^{(k)}$ defined through \eqref{eq:Tn-1}
and the Gelfand-Dikii polynomials $R^{(k)}$, so that they read
\begin{eqnarray}
\dot{\mathcal{J}} & = & \mathcal{D}^{(\mathcal{P})}T^{(k)}+\left(\mathcal{D}^{(\mathcal{J})}+a\mathcal{D}^{\left(\mathcal{P}\right)}\right)R^{\left(k\right)}\,,\nonumber \\
\dot{\mathcal{P}} & = & \mathcal{D}^{\left(\mathcal{P}\right)}R^{\left(k\right)}\,.\label{eq:EoMJPk}
\end{eqnarray}

It is worth pointing out that for any representative of the hierarchy
with $k>1$, not only the field equations in \eqref{eq:EoMJPk}, but
also the consistency condition for the functions $\varepsilon_{\mathcal{J}}$
and $\varepsilon_{\mathcal{P}}$ that parametrize the symmetries in
\eqref{eq:epsilon-punto} become severely more complicated than the
simplest cases of $k=0,1$ (see eqs. \eqref{eq:EoMJP0}, \eqref{eq:epsilonJP0},
and \eqref{eq:EoMJP1}, \eqref{eq:epsilonJP1}, respectively). However,
and remarkably, when $\varepsilon_{\mathcal{J}}$ and $\varepsilon_{\mathcal{P}}$
are assumed to be given by local functions of the dynamical fields
and their spatial derivatives, the general solution of the consistency
condition for the parameters in \eqref{eq:epsilon-punto} for $k\geq1$
turns out to be precisely given by the same expansion in terms of
the generalized Gelfand-Dikii polynomials $K^{(j)}$ and $\tilde{K}^{(j)}$,
as in \eqref{eq:sum-polynomials} for $k=1$. Therefore, as a consequence,
the corresponding canonical generators turn out to be given by the
two independent sets of conserved charges given by \eqref{eq:Q-1},
which are in involution, i.e., the commuting charges fulfill \eqref{eq:involution}
for both Poisson structures. Nevertheless, depending on the choice
of Poisson structure, $\mathcal{D}^{(2)}$ or $\mathcal{D}^{(1)}$,
the energy of the system in \eqref{eq:Energy}, now corresponds to
the Hamiltonian, $H^{(k)}$ or $H^{(k+1)}$, defined through \eqref{eq:Hk},
respectively.

As an ending remark of this subsection, it is worth noting that the
field equations of the so-called ``perturbed KdV hierarchy'' (see
e.g., \cite{0305-4470-16-16-015,MA199649}) are precisely recovered
from \eqref{eq:EoMJPk} for the particular case of $c_{\mathcal{J}}=a=0$.
In this case, for the entire hierarchy, according to \eqref{eq:Hk},
configurations with ${\cal J}=0$ do not carry energy, which goes
by hand with the fact that the Hamiltonian corresponds to the energy
of a perturbation described by ${\cal J}$. Indeed, for this class
of configurations, the entire set of conserved charges $\tilde{H}^{(j)}$
in \eqref{eq:Htilde} also vanishes. Note that the remaining ones,
given by $H_{\text{KdV}}^{\left(j\right)}$ generically remain nontrivial,
which can be interpreted as the conserved charges associated to an
arbitrary ``background configuration'' described by ${\cal P}={\cal P}\left(t,\phi\right)$
that solves the field equations of the $k$-th representative of the
KdV hierarchy.

\subsection{Analytic solutions\label{subsec:Analytic-solutions}}

Exact analytic solutions of the KdV equation, as well as for the $k$-th
representative of the KdV hierarchy, have been thoroughly studied
in the literature since long ago through different methods (see e.g.,
\cite{das1989integrable,Drazin:1989qi}). 

Here we show how to obtain an interesting wide class of analytic solutions
of the field equations in \eqref{eq:EoMJPk} for an arbitrary value
of the nonnegative integer $k$ that labels our hierarchy of integrable
systems with BMS$_{3}$ Poisson structure. 

As a warming up exercise, let us begin considering the simplest case,
described by choosing the field $\mathcal{P}$ to be constant, so
that the field equation of the KdV hierarchy in \eqref{eq:EoMJPk}
is trivially solved for an arbitrary value of $k$. In this case,
the field equation for $\mathcal{J}(t,\phi)$ in \eqref{eq:EoMJPk}
just reduces to a dispersive linear homogeneous equation with constant
coefficients, given by
\begin{equation}
\dot{\mathcal{J}}=\sum_{m=0}^{k}\alpha_{k,m}(-c_{\mathcal{P}})^{k-m}\mathcal{P}^{m}\partial_{\phi}^{2k-2m+1}\mathcal{J}\,,
\end{equation}
with $\alpha_{k,m}\equiv\frac{\left(2k+1\right)!!}{m!\left(2k-2m+1\right)!!}$,
which can be easily solved for an arbitrary member of the hierarchy.
Indeed, expanding in Fourier modes according to\footnote{$\mathcal{J}(t,\phi)$ is real provided that the modes fulfill $\left(\mathcal{J}_{n}\right)^{*}=\mathcal{J}_{-n}$.}

\begin{equation}
\mathcal{J}\left(t,\phi\right)=\frac{1}{2\pi}\sum_{n=0}^{\infty}{\cal J}_{n}e^{-i\left(\omega_{k,n}t+n\phi\right)}\,,\label{eq:JsolPcte}
\end{equation}
one finds that the corresponding dispersion relation is given by
\begin{equation}
\omega_{k,n}=\sum_{m=0}^{k}\alpha_{k,m}c_{\mathcal{P}}{}^{k-m}\mathcal{P}^{m}n{}^{2\left(k-m\right)+1}\,.
\end{equation}

In the case of nontrivial solutions $\mathcal{P}=\mathcal{P}(t,\phi)$
of the $k$-th KdV equation, it is also possible to find generic analytic
solutions for $\mathcal{J}\left(t,\phi\right)$. Remarkably, in spite
of the fact that $\mathcal{J}\left(t,\phi\right)$ obeys a linear
differential equation, their exact solutions are able to be nondispersive.
This effect occurs because the coefficients of the linear equation
for $\mathcal{J}$ in \eqref{eq:EoMJPk} are determined by nontrivial
solutions of the $k$-th KdV equation, and it persists even in presence
of a source term (with $a\neq0$ or $c_{\mathcal{J}}\neq0$). This
is explicitly discussed in what follows.

\subsubsection{Generic analytic solutions}

Let us assume that $\mathcal{P}=\mathcal{P}(t,\phi)$ corresponds
to an arbitrary generic solution of the field equations of the $k$-th
representative of the KdV hierarchy, described by the second line
of \eqref{eq:EoMJPk}. Note that, since the central extension $c_{\mathcal{P}}$
has not been scaled away, nontrivial solutions $\mathcal{P}(t,\phi)$
explicitly depend on $c_{\mathcal{P}}$. 

In order to find the form of $\mathcal{J}=\mathcal{J}(t,\phi)$ one
has to solve the remaining equation in \eqref{eq:EoMJPk}, which turns
out to be linear in $\mathcal{J}$, and possesses an inhomogeneous
term that is completely specified by $\mathcal{P}$ and their spatial
derivatives. Hence, the generic solution for $\mathcal{J}$ is given
by the sum of the particular and the homogeneous solutions, i.e.,
\begin{equation}
\mathcal{J}=\mathcal{J}_{h}+\mathcal{J}_{p}\,.
\end{equation}
Noteworthy, as it is shown in appendix \ref{sec:Proof}, for an arbitrary
value of the label $k$ of the hierarchy, a particular solution $\mathcal{J}=\mathcal{J}_{p}(t,\phi)$
can be analytically expressed in a very compact way, so that it reads\footnote{Note that this particular solution becomes trivial ($\mathcal{J}_{p}=0$)
in the case of ``perturbed KdV'' described by $c_{\mathcal{J}}=a=0$. }
\begin{equation}
\mathcal{J}_{p}=c_{\mathcal{J}}\frac{\partial\mathcal{P}}{\partial c_{\mathcal{P}}}+at\dot{\mathcal{P}}\,.\label{eq:Jparticular}
\end{equation}
Besides, a generic solution of the homogeneous equation can be found
by virtue of the symmetries in \eqref{eq:delta-P}, being spanned
by the subset of parameters given by \eqref{eq:sum-polynomials} that
preserve the form of $\mathcal{P}(t,\phi)$, i.e., the ones for which
$\delta\mathcal{P}=0$. The suitable subset of symmetries we look
for then becomes generated by an arbitrary combination of the generalized
polynomials $K^{(j)}$, excluding $\tilde{K}^{(j)}$; and hence the
parameters are given by $\varepsilon_{\mathcal{J}}$ and $\varepsilon_{\mathcal{P}}$
in \eqref{eq:sum-polynomials} with $\tilde{\eta}_{j}=0$. Thus, according
to \eqref{eq:delta-P}, the homogeneous solution is given by
\begin{equation}
\mathcal{J}_{h}=\delta_{\eta_{j}}\mathcal{J}=\sum_{j=0}^{\infty}\eta_{j}\mathcal{D}^{(\mathcal{P})}R^{(j)}\,.\label{eq:Jhomogenea}
\end{equation}
Therefore, by virtue of the recursive relation of the Gelfand-Dikii
polynomials in \eqref{eq:Gelfan-1}, the generic solution for $\mathcal{J}$
acquires the form
\begin{equation}
\mathcal{J}=\sum_{j=0}^{\infty}\eta_{j}\partial_{\phi}R^{(j+1)}+c_{\mathcal{J}}\frac{\partial\mathcal{P}}{\partial c_{\mathcal{P}}}+at\dot{\mathcal{P}}\,.\label{eq:Jgeneral}
\end{equation}

As a cross-check, in appendix \ref{sec:Proof} it is explicitly shown
that eq. \eqref{eq:Jgeneral} solves the field equation for $\mathcal{J}$
in \eqref{eq:EoMJPk}.

In sum, the generic solution for $\mathcal{J}$ in \eqref{eq:Jgeneral}
has been generated through acting on the particular solution $\mathcal{J}_{p}$
with the symmetries that are spanned by the corresponding canonical
generators in \eqref{eq:Q-1} with $\tilde{\eta}_{j}=0$. Hence, since
the generators are in involution, the full set of conserved charges
for the solution characterized by the fields $\mathcal{P}$ and $\mathcal{J}_{p}$
must coincide with the ones for the fields $\mathcal{P}$ with $\mathcal{J}$
given by \eqref{eq:Jgeneral}. In other words, the homogeneous part
of the solution $\mathcal{J}_{h}$ does not contribute to the conserved
charges. 

The conserved charges are then described by the corresponding ones
for KdV, given by $H_{\text{KdV}}^{\left(n\right)}$, together with
the independent set $\tilde{H}^{\left(n\right)}$ in \eqref{eq:Htilde}.
Once the latter set is evaluated in the generic solution \eqref{eq:Jgeneral},
it can be compactly written in terms of the total derivative of $H_{\text{KdV}}^{\left(n\right)}$
with respect to $c_{\mathcal{P}}$, so that it reads (see appendix
\ref{sec:Proof})
\begin{equation}
\tilde{H}^{(n)}=c_{\mathcal{J}}\frac{dH_{\text{KdV}}^{(n)}}{dc_{\mathcal{P}}}\,.\label{eq:ConservedSolutions}
\end{equation}

Therefore, for the BMS$_{3}$ Poisson structure $\mathcal{D}^{\left(2\right)}$,
the energy of our generic solution for the $k$-th representative
of the hierarchy is given by the Hamiltonian in \eqref{eq:Hk}, which
reduces to
\begin{equation}
H^{(k)}=c_{\mathcal{J}}\frac{dH_{\text{KdV}}^{(k)}}{dc_{\mathcal{P}}}+aH_{\text{KdV}}^{(k)}\,.\label{eq:EnergySolutions}
\end{equation}

It is worth pointing out that the conserved charges of the solution
can be expressed exclusively in terms of $\mathcal{P}$ and their
spatial derivatives. 

Note that the generic class of solutions presented here is mapped
into itself under the anisotropic scaling of Lifshitz type given in
\eqref{eq:Lifshitz-z-scaling}, where the arbitrary constants $\eta_{j}$
of the homogeneous solution \eqref{eq:Jhomogenea} transform as $\bar{\eta}_{j}=\sigma^{2j-1}\eta_{j}$.

Additionally, as pointed in section \ref{RemarksSymm}, in the case
of $k=1$ the field equations are also invariant under Galilean transformations.
Hence, in this case, by virtue of the Galilean boost spanned by \eqref{eq:Galilean-boost},
our solution in \eqref{eq:Jgeneral} acquires nontrivial zero modes
once expressed in the moving frame.

In the next subsection, we explicitly describe a couple of simple
and interesting particular examples of analytic solutions in the case
of $k=1$.

\subsubsection{Particular cases for $k=1$\label{subsec:Particular-cases}}

\emph{Single KdV soliton on the real line.-} The integrable system
described in section \ref{subsec:Integrable-bi-Hamiltonian-systemk=00003D1}
can be extended to $\mathbb{R}^{2}$ provided that the angular coordinate
is unwrapped $(-\infty<\phi<\infty)$, so that our previous analysis
still holds once the fall-off of the fields is assumed to be fast
enough so as to get rid of boundary terms. 

In our conventions, the well-known single soliton solution of the
KdV equation in \eqref{eq:EoMJP1} reads
\begin{equation}
\mathcal{P}=-v\,\text{sech}^{2}\left(x\right)\,,\label{eq:Psoliton}
\end{equation}
where $x=\sqrt{\frac{v}{4c_{\mathcal{P}}}}(\phi-vt)$, and $v$ stands
for the integration constant that parametrizes the velocity and amplitude
of the soliton. 

\begin{figure}[h]\begin{center}\begin{tikzpicture}[scale=0.7]   \draw (-5.999997, -0.018389678422198893) -- (-5.909827976440004, -0.020606703176161523) -- (-5.819658952880008, -0.02308450981979038) -- (-5.729489929320011, -0.025852694435264292) -- (-5.639320905760016, -0.028944024553105692) -- (-5.549151882200021, -0.0323947443267999) -- (-5.458982858640025, -0.03624490206575436) -- (-5.368813835080028, -0.04053870014286204) -- (-5.278644811520032, -0.04532486679217895) -- (-5.188475787960037, -0.05065704865533413) -- (-5.09830676440004, -0.05659422208520659) -- (-5.008137740840044, -0.06320112012729998) -- (-4.917968717280049, -0.07054867071911343) -- (-4.827799693720054, -0.07871443991081484) -- (-4.737630670160057, -0.08778307173926952) -- (-4.6474616466000604, -0.09784671368999368) -- (-4.557292623040065, -0.10900541334936266) -- (-4.467123599480069, -0.12136746775530498) -- (-4.376954575920073, -0.1350497019512548) -- (-4.286785552360077, -0.15017764716614163) -- (-4.196616528800081, -0.16688558169141754) -- (-4.1064475052400855, -0.18531638869240324) -- (-4.016278481680089, -0.2056211746462215) -- (-3.9261094581200933, -0.22795857960313218) -- (-3.835940434560097, -0.25249369578630165) -- (-3.7457714110001015, -0.2793964939654053) -- (-3.6556023874401053, -0.3088396374071954) -- (-3.5654333638801097, -0.340995540971491) -- (-3.4752643403201136, -0.37603250820848594) -- (-3.385095316760118, -0.4141097525196126) -- (-3.2949262932001218, -0.4553710803729971) -- (-3.11458824608013, -0.5478948862332277) -- (-3.016834691990136, -0.6037665860298664) -- (-2.9190811379001422, -0.6636479714638) -- (-2.7235740297201545, -0.7947656965649206) -- (-2.62582047563016, -0.8652817743602316) -- (-2.5280669215401663, -0.9382896263175368) -- (-2.3325598133601786, -1.0877629474194612) -- (-2.2348062592701847, -1.1613863848855372) -- (-2.137052705180191, -1.2317150653226197) -- (-2.039299151090197, -1.2962714690866561) -- (-1.941545597000203, -1.3520874738368711) -- (-1.843792042910209, -1.395695293619761) -- (-1.7460384888202152, -1.4231522264043817) -- (-1.648284934730221, -1.4301129924917189) -- (-1.5505313806402272, -1.411964837834112) -- (-1.4527778265502334, -1.3640403364021612) -- (-1.3550242724602393, -1.2819198906689089) -- (-1.2572707183702454, -1.1618291120041673) -- (-1.1595171642802515, -1.0011245105026967) -- (-1.0617636101902577, -0.7988437542178846) -- (-0.9640100561002638, -0.5562748573546847) -- (-0.8662565020102699, -0.2774745274938552) -- (-0.768502947920276,    0.030355692479054314) -- (-0.670749393830282,    0.3567390120395776) -- (-0.5729958397402881,    0.6881370256233049) -- (-0.47524228565029425,    1.0085764331668143) -- (-0.37748873156030033,    1.3007022644168955) -- (-0.27973517747030635,    1.5471942164677097) -- (-0.18198162338031248,    1.7323987655693294) -- (-0.08422806929031856,    1.8439596987349365) -- (0.013525484799675346,    1.874196160771207) -- (0.10480107645091723,    1.8270594126216466) -- (0.19607666810215912,    1.709992668186988) -- (0.287352259753401,    1.5300033530459167) -- (0.3786278514046429,    1.2975282366945424) -- (0.4699034430558847,    1.0254221151869491) -- (0.5611790347071266,    0.7277918603901099) -- (0.7437302180096105,    0.11179831176522757) -- (0.8350058096608524, -0.18178581623668075) -- (0.9262814013120941, -0.45264124172771314) -- (1.017556992963336, -0.6939642483374786) -- (1.108832584614578, -0.9014110108329271) -- (1.2001081762658197, -1.072892632140531) -- (1.2913837679170617, -1.2082454729893757) -- (1.3826593595683037, -1.308838299951673) -- (1.4739349512195454, -1.3771673748244637) -- (1.5652105428707872, -1.4164769503485646) -- (1.6564861345220294, -1.4304288046553477) -- (1.7477617261732712, -1.4228324251318505) -- (1.8390373178245127, -1.397438197475952) -- (1.930312909475755, -1.3577896394647353) -- (2.0215885011269967, -1.3071270218291853) -- (2.1128640927782385, -1.2483330861141342) -- (2.2041396844294803, -1.1839113970284412) -- (2.2954152760807225, -1.115988610874057) -- (2.3866908677319643, -1.0463331700971465) -- (2.5692420510344482, -0.9072868513913805) -- (2.6605176426856896, -0.8399277431548029) -- (2.751793234336932, -0.7749728441873649) -- (2.934344417639416, -0.6540378691459823) -- (3.030146060838184, -0.5959209842516277) -- (3.1259477040369523, -0.5416644558722432) -- (3.2217493472357206, -0.491281409421393) -- (3.3175509904344884, -0.44471170247902225) -- (3.4133526336332563, -0.40184086791733864) -- (3.509154276832025, -0.3625155955906484) -- (3.6049559200307932, -0.326556216433978) -- (3.7007575632295615, -0.293766632789241) -- (3.79655920642833, -0.26394210308374794) -- (3.892360849627098, -0.23687524562554757) -- (3.9881624928258663, -0.2123605805744354) -- (4.083964136024635, -0.19019788446380975) -- (4.179765779223403, -0.1701945899706856) -- (4.275567422422172, -0.15216742599210464) -- (4.37136906562094, -0.13594345988753165) -- (4.467170708819708, -0.12136067498872044) -- (4.562972352018476, -0.10826819192082644) -- (4.658773995217244, -0.09652622155110524) -- (4.7545756384160125, -0.0860058200572602) -- (4.85037728161478, -0.07658850224888486) -- (4.946178924813549, -0.06816575746282322) -- (5.041980568012318, -0.06063850270037117) -- (5.137782211211086, -0.05391649983563767) -- (5.233583854409854, -0.04791775739748213) -- (5.329385497608622, -0.04256793235139041) -- (5.425187140807391, -0.03779974326029501) -- (5.5209887840061596, -0.03355240299740188) -- (5.616790427204927, -0.029771076662922228) -- (5.712592070403696, -0.026406368390115665) -- (5.808393713602464, -0.023413839206781816) -- (5.904195356801232, -0.02075355695799324) -- (5.999997, -0.018389678422198893);
\draw [densely dotted,line width=0.8] (-5.999997, -0.02185409592895294) -- (-5.909827976440004, -0.024552282996394934) -- (-5.819658952880008, -0.027578018927451924) -- (-5.729489929320011, -0.030970146625348905) -- (-5.639320905760016, -0.03477195575222553) -- (-5.549151882200021, -0.03903166221062035) -- (-5.458982858640025, -0.04380293333515886) -- (-5.368813835080028, -0.04914546179744569) -- (-5.278644811520032, -0.05512559106519399) -- (-5.188475787960037, -0.06181699496322965) -- (-5.09830676440004, -0.06930141341905968) -- (-5.008137740840044, -0.07766944578901122) -- (-4.917968717280049, -0.08702140219004158) -- (-4.827799693720054, -0.09746821192965958) -- (-4.737630670160057, -0.10913238633625866) -- (-4.6474616466000604, -0.12214903092677987) -- (-4.557292623040065, -0.13666689876389626) -- (-4.467123599480069, -0.15284947287537426) -- (-4.376954575920073, -0.17087606052153506) -- (-4.286785552360077, -0.19094287564748055) -- (-4.196616528800081, -0.21326407774043418) -- (-4.1064475052400855, -0.23807272516937591) -- (-4.016278481680089, -0.2656215884939806) -- (-3.9261094581200933, -0.2961837537090794) -- (-3.835940434560097, -0.33005292639308464) -- (-3.7457714110001015, -0.3675433246518664) -- (-3.6556023874401053, -0.4089890209522517) -- (-3.5654333638801097, -0.4547425597733977) -- (-3.4752643403201136, -0.5051726388681728) -- (-3.385095316760118, -0.5606605963510912) -- (-3.2949262932001218, -0.6215953936067169) -- (-3.2047572696401256, -0.6883667253866246) -- (-3.11458824608013, -0.7613558243990487) -- (-3.016834691990136, -0.8479268398715054) -- (-2.9190811379001422, -0.9426400805231877) -- (-2.821327583810148, -1.0458276063493153) -- (-2.7235740297201545, -1.157705475357744) -- (-2.62582047563016, -1.2783278345473315) -- (-2.5280669215401663, -1.4075309790438095) -- (-2.3325598133601786, -1.6895229870814759) -- (-2.2348062592701847, -1.8402370213056196) -- (-2.137052705180191, -1.995194532527944) -- (-2.039299151090197, -2.1519265691206173) -- (-1.941545597000203, -2.3072051093751087) -- (-1.843792042910209, -2.456948243762164) -- (-1.7460384888202152, -2.596147869559217) -- (-1.648284934730221, -2.718837492738139) -- (-1.5505313806402272, -2.8181221839610537) -- (-1.4527778265502334, -2.886296357059177) -- (-1.3550242724602393, -2.9150765443445925) -- (-1.2572707183702454, -2.8959740156922913) -- (-1.1595171642802515, -2.820823938936763) -- (-1.0617636101902577, -2.682471948264928) -- (-0.9640100561002638, -2.47559443714825) -- (-0.8662565020102699, -2.1975963578046405) -- (-0.768502947920276, -1.8494933093350001) -- (-0.670749393830282, -1.4366501955297135) -- (-0.5729958397402881, -0.9692268628689521) -- (-0.37748873156030033, 0.06526626626464221) -- (-0.27973517747030635,    0.5907917860666425) -- (-0.18198162338031248,    1.0906125277737182) -- (-0.08422806929031856,    1.54152050662966) -- (0.013525484799675346,    1.922995850247125) -- (0.10480107645091723,    2.2023573017491134) -- (0.19607666810215912,   2.3989503825230027) -- (0.287352259753401,    2.509550008856509) -- (0.3786278514046429,    2.53597098335168) -- (0.4699034430558847,    2.4845561150411695) -- (0.5611790347071266,    2.365277395494962) -- (0.6524546263583685,    2.190600235658901) -- (0.7437302180096105,    1.9742745143155793) -- (0.8350058096608524,    1.730200096332795) -- (0.9262814013120941,    1.4714771926176897) -- (1.017556992963336,    1.2097050002645293) -- (1.108832584614578,    0.9545463478509758) -- (1.2001081762658197,    0.7135393212100951) -- (1.2913837679170617,    0.4921127880061006) -- (1.3826593595683037,    0.29375151447583375) -- (1.4739349512195454,    0.12025589476750892) -- (1.5652105428707872, -0.027952183261273356) -- (1.6564861345220294, -0.15151553017653058) -- (1.7477617261732712, -0.2518468041578384) -- (1.8390373178245127, -0.33086908364179296) -- (1.930312909475755, -0.3907997776235257) -- (2.0215885011269967, -0.43397945954271033) -- (2.1128640927782385, -0.46274300336433716) -- (2.2041396844294803, -0.47932797106709824) -- (2.2954152760807225, -0.4858141205254064) -- (2.3866908677319643, -0.4840877555023785) -- (2.477966459383206, -0.47582507276395014) -- (2.5692420510344482, -0.46248940465062977) -- (2.6605176426856896, -0.4453381180161408) -- (2.751793234336932, -0.42543578900703766) -- (2.8430688259881736, -0.40367105644351287) -- (2.934344417639416, -0.38077523028209925) -- (3.030146060838184, -0.356174143148587) -- (3.1259477040369523, -0.3315232240353527) -- (3.3175509904344884, -0.28365072433739763) -- (3.4133526336332563, -0.2609776480065212) -- (3.509154276832025, -0.23938516233206822) -- (3.7007575632295615, -0.19981559568311796) -- (3.79655920642833, -0.18192203001643364) -- (3.892360849627098, -0.1652937826796274) -- (3.9881624928258663, -0.14990652992939527) -- (4.083964136024635, -0.13572055970407626) -- (4.179765779223403, -0.12268518825578606) -- (4.275567422422172, -0.11074230836807426) -- (4.37136906562094, -0.09982921011234072) -- (4.467170708819708, -0.08988079690920042) -- (4.562972352018476, -0.08083130259747245) -- (4.658773995217244, -0.07261559965154112) -- (4.7545756384160125, -0.06517017478772566) -- (4.85037728161478, -0.05843383598063774) -- (4.946178924813549, -0.05234820430233287) -- (5.041980568012318, -0.0468580348811825) -- (5.137782211211086, -0.04191140350979277) -- (5.233583854409854, -0.03745978885977868) -- (5.329385497608622, -0.03345807473596593) -- (5.425187140807391, -0.029864492182471863) -- (5.5209887840061596, -0.026640517409007945) -- (5.616790427204927, -0.023750738322034212) -- (5.712592070403696, -0.02116269982019827) -- (5.808393713602464, -0.01884673585812317) -- (5.904195356801232, -0.01677579452062363) -- (5.999997, -0.01492526091544485);
\draw [dashed] (-5.999997, -0.027050722189084008) -- (-5.909827976440004, -0.030470652726745043) -- (-5.819658952880008, -0.03431828258894424) -- (-5.729489929320011, -0.03864632491047583) -- (-5.639320905760016, -0.043513852550905285) -- (-5.549151882200021, -0.048987039036351) -- (-5.458982858640025, -0.05513998023926562) -- (-5.368813835080028, -0.06205560427932118) -- (-5.278644811520032, -0.06982667747471658) -- (-5.188475787960037, -0.07855691442507293) -- (-5.09830676440004, -0.0883622004198393) -- (-5.008137740840044, -0.09937193428157812) -- (-4.917968717280049, -0.1117304993964338) -- (-4.827799693720054, -0.12559886995792668) -- (-4.737630670160057, -0.14115635823174236) -- (-4.6474616466000604, -0.15860250678195914) -- (-4.557292623040065, -0.17815912688569666) -- (-4.467123599480069, -0.2000724805554781) -- (-4.376954575920073, -0.2246155983769554) -- (-4.286785552360077, -0.252090718369489) -- (-4.196616528800081, -0.2828318218139591) -- (-4.1064475052400855, -0.31720722988483485) -- (-4.016278481680089, -0.35562220926561916) -- (-3.9261094581200933, -0.3985215148680002) -- (-3.835940434560097, -0.4463917723032592) -- (-3.7457714110001015, -0.49976357068155797) -- (-3.6556023874401053, -0.5592130962698361) -- (-3.5654333638801097, -0.6253630879762577) -- (-3.4752643403201136, -0.6988828348577028) -- (-3.385095316760118, -0.7804868620983091) -- (-3.2949262932001218, -0.8709318634572965) -- (-3.2047572696401256, -0.9710113335555046) -- (-3.11458824608013, -1.08154723164778) -- (-3.016834691990136, -1.2141672206339635) -- (-2.9190811379001422, -1.361128244112269) -- (-2.821327583810148, -1.5234611674460217) -- (-2.7235740297201545, -1.7021151435469792) -- (-2.62582047563016, -1.8978969248279816) -- (-2.5280669215401663, -2.111393008133219) -- (-2.3325598133601786, -2.5921630465744983) -- (-2.2348062592701847, -2.8585129759357426) -- (-2.137052705180191, -3.1404137333359303) -- (-1.941545597000203, -3.7398815626824655) -- (-1.843792042910209, -4.048827668975767) -- (-1.7460384888202152, -4.3556413342914695) -- (-1.648284934730221, -4.651924243107768) -- (-1.5505313806402272, -4.927358203151466) -- (-1.4527778265502334, -5.169680388044701) -- (-1.3550242724602393, -5.364811524858117) -- (-1.2572707183702454, -5.497191371224478) -- (-1.1595171642802515, -5.5503730815878605) -- (-1.0617636101902577, -5.507914239335494) -- (-0.9640100561002638, -5.3545738068386) -- (-0.8662565020102699, -5.077779103270819) -- (-0.768502947920276, -4.669266812056082) -- (-0.670749393830282, -4.126734006883649) -- (-0.5729958397402881, -3.455272695607338) -- (-0.47524228565029425, -2.668323173640401) -- (-0.37748873156030033, -1.7878877309637378) -- (-0.27973517747030635, -0.8438118595349581) -- (-0.18198162338031248, 0.1279331710803014) -- (-0.08422806929031856,    1.087861718471745) -- (0.013525484799675346,    1.9961953844610023) -- (0.10480107645091723,    2.7653041354403127) -- (0.19607666810215912,    3.4323869540270255) -- (0.287352259753401,    3.978869992572397) -- (0.3786278514046429,    4.3936351033373855) -- (0.4699034430558847,    4.6732571148224995) -- (0.5611790347071266,    4.821505698152241) -- (0.6524546263583685,   4.848252507404204) -- (0.7437302180096105,    4.767988818141106) -- (0.8350058096608524,    4.598178965187008) -- (0.9262814013120941,    4.357654844135794) -- (1.017556992963336,    4.065208873167541) -- (1.108832584614578,    3.7384823858768303) -- (1.2001081762658197,    3.393187251236035) -- (1.2913837679170617,    3.042650179499315) -- (1.4739349512195454,    2.3663907991554676) -- (1.5652105428707872,    2.0548349673696635) -- (1.6564861345220294,    1.7668543815416955) -- (1.7477617261732712,    1.50463162730318) -- (1.8390373178245127,    1.2689845871094456) -- (1.930312909475755,    1.0596850151382886) -- (2.0215885011269967,    0.8757418838870024) -- (2.1128640927782385,    0.7156421207603583) -- (2.2041396844294803,    0.5775471678749164) -- (2.2954152760807225,    0.4594476149975695) -- (2.3866908677319643,    0.3592803663897738) -- (2.477966459383206,    0.27501382290634546) -- (2.5692420510344482,    0.2047067654604963) -- (2.6605176426856896,    0.1465463196918522) -- (2.751793234336932,    0.09886979376345306) -- (2.8430688259881736,    0.06017447264531184) -- (2.934344417639416,    0.029118728013725198) -- (3.030146060838184,    0.003446118505974066) -- (3.1259477040369523, -0.01631137628001718) -- (3.2217493472357206, -0.031185523406081356) -- (3.3175509904344884, -0.042059257124960614) -- (3.4133526336332563, -0.049682818140295015) -- (3.509154276832025, -0.05468951244419793) -- (3.6049559200307932, -0.05761060158820039) -- (3.7007575632295615, -0.05888904002393346) -- (3.79655920642833, -0.05889192041546226) -- (3.892360849627098, -0.05792158826074714) -- (3.9881624928258663, -0.05622545396183501) -- (4.083964136024635, -0.05400457256447597) -- (4.179765779223403, -0.05142108568343676) -- (4.275567422422172, -0.04860463193202869) -- (4.467170708819708, -0.04266097978992037) -- (4.562972352018476, -0.03967596861244148) -- (4.658773995217244, -0.0367496668021949) -- (4.85037728161478, -0.031201836578267077) -- (4.946178924813549, -0.028621874561597337) -- (5.041980568012318, -0.02618733315239951) -- (5.233583854409854, -0.021772836053223513) -- (5.329385497608622, -0.019793288312829206) -- (5.425187140807391, -0.017961615565737134) -- (5.5209887840061596, -0.01627268902641703) -- (5.616790427204927, -0.014720230810702187) -- (5.712592070403696, -0.013297196965322184) -- (5.808393713602464, -0.0119960808351352) -- (5.904195356801232, -0.010809150864569213) -- (5.999997, -0.009728634655313782);
    \draw [latex-latex,line width=1] (0,-6) -- (0,5.5);     \draw [latex-latex,line width=1] (-6.5,0) -- (6.5,0);
    \draw (6.3,-0.3) node {} node {$x$};     \draw (-0.3,5.3) node {} node {$\mathcal{J}$};  
\end{tikzpicture}\end{center}
\caption{The form of $\mathcal{J}$ in \eqref{eq:SolJamig} is plotted for generic fixed values of $c_\mathcal{P}$, $c_\mathcal{J}$ , $a$, $v$ and for different values of $\eta_1$. Note that when the integration constant $\eta_1$ vanishes,  $\mathcal{J}$ becomes an even function of $x$ (solid line).}
\label{Figure_1}
\end{figure}
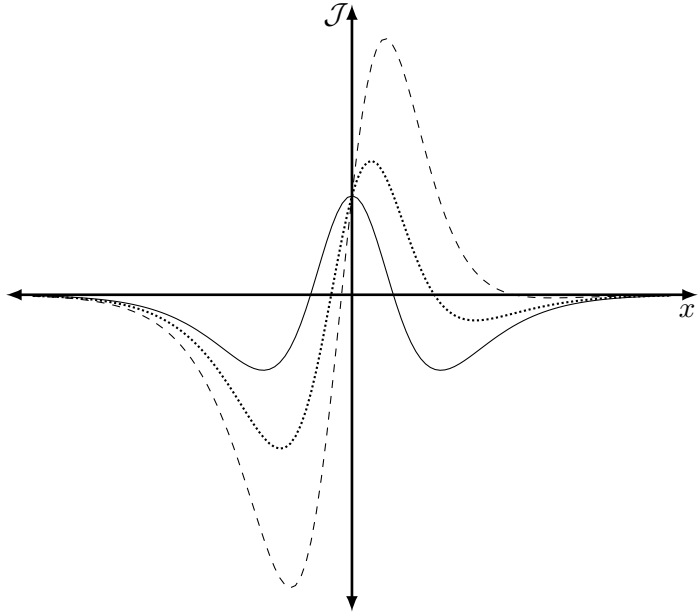

An analytic solution for the remaining field equation in \eqref{eq:EoMJP1}
can then be constructed out from \eqref{eq:Jgeneral}, with $\mathcal{P}$
given by \eqref{eq:Psoliton}. For simplicity we consider that the
integration constants in \eqref{eq:Jgeneral} are chosen as $\eta_{j}=\eta_{1}\delta_{j,1}$,
with $\eta_{1}$ arbitrary, so that the solution for $\mathcal{J}$
becomes explicitly given by
\begin{eqnarray}
\mathcal{J} & = & \left(\eta_{1}\frac{v^{3/2}}{\sqrt{c_{\mathcal{P}}}}-\frac{vx(ac_{\mathcal{P}}+c_{\mathcal{J}})}{c_{\mathcal{P}}}\right)\tanh\left(x\right)\text{sech}^{2}\left(x\right)+av\,\text{sech}^{2}\left(x\right)\,.\label{eq:SolJamig}
\end{eqnarray}
Therefore, although $\mathcal{J}$ obeys a linear differential equation,
the solution in \eqref{eq:SolJamig} clearly maintains its shape as
it evolves in time. The profile of $\mathcal{J}$ in \eqref{eq:SolJamig}
is depicted in figure \ref{Figure_1}.

\medskip{}

\emph{Cnoidal wave on $S^{1}$.-} In the case of periodic boundary
conditions $(-\pi\leq\phi<\pi)$ an analytic solution for the KdV
equation in \eqref{eq:EoMJP1} of solitonic type is known as a ``cnoidal
wave'', since it is described in terms the Jacobi elliptic cosine
(cn) (see e.g. \cite{Drazin:1989qi}). The solution is given by
\begin{equation}
\mathcal{P}=4c_{\mathcal{P}}\left[A-\alpha\text{cn}^{2}\left(y,m\right)\right]\,,\label{eq:Pcnoidal}
\end{equation}
with $y=\sqrt{\frac{\alpha}{m}}(\phi-c_{\mathcal{P}}vt)$, and the
velocity parameter is related to the remaining integration constants
as
\begin{equation}
v=4\alpha\left(2-\frac{1}{m}\right)-12A\,.\label{eq:Constrain}
\end{equation}
The wavelength of the solution is given by $2\sqrt{\frac{m}{\alpha}}K(m)$,
where $K$ stands for a complete elliptic integral of the first kind.
Accordingly, the elliptic parameter $m$ can take values within the
range $0<m<1$, satisfying $2\sqrt{\frac{m}{\alpha}}K(m)=\frac{2\pi}{n}$
with $n\in\mathbb{N}$. 

As in the previous example, we then construct the analytic solution
for $\mathcal{J}$ from the generic one in \eqref{eq:Jgeneral}, with
$\mathcal{P}$ given by the cnoidal wave in \eqref{eq:Pcnoidal}.
For the sake of simplicity, we again choose the integration constants
to be given by $\eta_{j}=\eta_{1}\delta_{j,1}$, and hence the searched
for analytic solution becomes
\begin{equation}
\mathcal{J}=4c_{\mathcal{J}}\left[A-\alpha\text{cn}^{2}\left(y,m\right)\right]+\left[8\alpha c_{\mathcal{P}}\sqrt{\frac{\alpha}{m}}(\eta_{1}-(ac_{\mathcal{P}}+c_{\mathcal{J}})vt)\right]\text{cn}\left(y,m\right)\text{sn}\left(y,m\right)\text{dn}\left(y,m\right)\,,\label{eq:J cnoidal}
\end{equation}
where sn and dn stand for the elliptic sine and the delta amplitude,
respectively. 

Note that the profile of $\mathcal{J}$ preserves its form as it evolves
in time only in the case of $\gamma=ac_{\mathcal{P}}+c_{\mathcal{J}}=0$
(perturbed KdV), otherwise the amplitude grows linearly with time.
Nonetheless, the energy in \eqref{eq:EnergySolutions} as well as
the remaining conserved charges given by $H_{\text{KdV}}^{\left(n\right)}$,
and $\tilde{H}^{\left(n\right)}$ in \eqref{eq:Htilde}, turn out
to be finite regardless the value of $\gamma$.

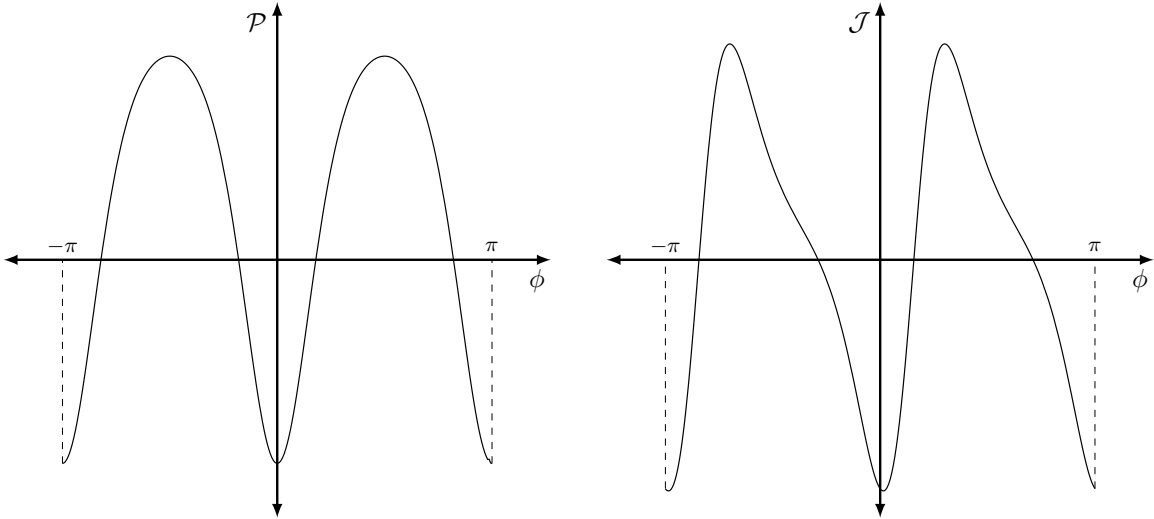
\begin{figure}
\begin{center}
\scalebox{0.9}{\begin{tikzpicture}[scale=1.0]   \draw [line width=0.5] 
(-3.14159, -3.) -- (-3.12061, -2.99382) -- (-3.09963, -2.97533) --  (-3.07864, -2.94467) -- (-3.05766, -2.90207) -- (-3.03668, -2.84785) --  (-3.01569, -2.78241) -- (-2.99471, -2.70623) -- (-2.97373, -2.61987) --  (-2.95274, -2.52391) -- (-2.93176, -2.41904) -- (-2.88979, -2.18537) --  (-2.86881, -2.05807) -- (-2.84783, -1.92483) -- (-2.80586, -1.64362) --  (-2.78488, -1.4972) -- (-2.76389, -1.34792) -- (-2.72193, -1.04362) --  (-2.63799, -0.430307) -- (-2.61701, -0.279333) -- (-2.59603, -0.130269) --  (-2.55406, 0.160551) -- (-2.47013, 0.704333) -- (-2.44738,    0.841545) -- (-2.42463, 0.974053) -- (-2.37913, 1.22452) -- (-2.35639,    1.34236) -- (-2.33364, 1.45523) -- (-2.28814, 1.66616) -- (-2.26539,    1.76432) -- (-2.24264, 1.8577) -- (-2.19715, 2.0305) -- (-2.1744,    2.11014) -- (-2.15165, 2.18543) -- (-2.10615, 2.32345) -- (-2.08341,    2.38645) -- (-2.06066, 2.44562) -- (-2.03791, 2.50109) -- (-2.01516,    2.55298) -- (-1.99241, 2.60143) -- (-1.96966, 2.64657) -- (-1.92417,    2.72735) -- (-1.90142, 2.76322) -- (-1.87867, 2.79622) -- (-1.85592,    2.82645) -- (-1.83317, 2.854) -- (-1.81043, 2.87896) -- (-1.78768,    2.9014) -- (-1.76493, 2.9214) -- (-1.74218, 2.93901) -- (-1.72094,    2.95337) -- (-1.6997, 2.96574) -- (-1.67846, 2.97617) -- (-1.65722,    2.98468) -- (-1.63598, 2.9913) -- (-1.61474, 2.99605) -- (-1.5935,    2.99895) -- (-1.57225, 3.) -- (-1.55101, 2.9992) -- (-1.52977,    2.99656) -- (-1.50853, 2.99206) -- (-1.48729, 2.9857) -- (-1.46605,    2.97745) -- (-1.44481, 2.96729) -- (-1.42357, 2.95518) -- (-1.40233,    2.9411) -- (-1.38109, 2.92501) -- (-1.35985, 2.90685) -- (-1.33861,    2.88657) -- (-1.31737, 2.86412) -- (-1.29612, 2.83944) -- (-1.27488,    2.81246) -- (-1.2324, 2.75129) -- (-1.21116, 2.71695) -- (-1.18992,    2.67999) -- (-1.16868, 2.64032) -- (-1.14744, 2.59784) -- (-1.1262,    2.55246) -- (-1.10496, 2.50408) -- (-1.06248, 2.39789) -- (-1.04165,    2.34105) -- (-1.02083, 2.2809) -- (-0.979179, 2.15034) -- (-0.958355,    2.07972) -- (-0.937531, 2.00541) -- (-0.895883, 1.84539) -- (-0.875058,    1.7595) -- (-0.854234, 1.66961) -- (-0.812586, 1.47759) -- (-0.729289,    1.04384) -- (-0.708465, 0.925112) -- (-0.687641, 0.802371) -- (-0.645993,    0.545301) -- (-0.562696, -0.0102814) -- (-0.396103, -1.2152) --  (-0.373514, -1.3779) -- (-0.350925, -1.53794) -- (-0.305747, -1.84639) --  (-0.283158, -1.99289) -- (-0.260568, -2.13295) -- (-0.21539, -2.38987) --  (-0.192801, -2.50483) -- (-0.170212, -2.60961) -- (-0.147623, -2.70336) --  (-0.125034, -2.78533) -- (-0.102444, -2.85483) -- (-0.0798553, -2.91127) --  (-0.0572661, -2.95416) -- (-0.034677, -2.98314) -- (-0.0120878,  -2.99795) -- (0.0105013, -2.99845) -- (0.0330905, -2.98465) -- (0.0556796,  -2.95666) -- (0.0782687, -2.91473) -- (0.100858, -2.85923) -- (0.123447,  -2.79063) -- (0.146036, -2.70951) -- (0.168625, -2.61656) -- (0.191214,  -2.51253) -- (0.213804, -2.39826) -- (0.236393, -2.27461) -- (0.258982,  -2.14252) -- (0.281571, -2.00295) -- (0.326749, -1.70525) -- (0.347831,  -1.5596) -- (0.368913, -1.41074) -- (0.411076, -1.10634) -- (0.495402,  -0.489712) -- (0.516484, -0.337405) -- (0.537566, -0.186856) -- (0.579729,    0.107301) -- (0.664056, 0.658643) -- (0.685137, 0.787348) -- (0.706219,    0.912063) -- (0.748382, 1.14912) -- (0.769464, 1.26133) -- (0.790546,    1.36929) -- (0.832709, 1.57242) -- (0.853791, 1.66765) -- (0.874872,    1.75872) -- (0.917036, 1.92859) -- (0.938117, 2.00756) -- (0.959199,    2.08265) -- (1.00136, 2.22162) -- (1.02421, 2.2909) -- (1.04706,    2.35611) -- (1.0699, 2.41739) -- (1.09275, 2.47488) -- (1.1156,    2.5287) -- (1.13844, 2.57898) -- (1.18414, 2.66946) -- (1.20698,    2.70989) -- (1.22983, 2.74727) -- (1.25268, 2.7817) -- (1.27552,    2.8133) -- (1.29837, 2.84216) -- (1.32122, 2.86836) -- (1.34406,    2.89198) -- (1.36691, 2.91312) -- (1.38976, 2.93182) -- (1.4126,    2.94816) -- (1.43545, 2.96219) -- (1.4583, 2.97396) -- (1.48114,    2.9835) -- (1.50399, 2.99086) -- (1.52684, 2.99605) -- (1.54968,    2.99909) -- (1.57253, 2.99999) -- (1.59538, 2.99877) -- (1.61822,    2.9954) -- (1.64107, 2.98988) -- (1.66392, 2.9822) -- (1.68676,    2.97232) -- (1.70961, 2.96021) -- (1.73245, 2.94583) -- (1.75488,    2.92946) -- (1.77731, 2.91081) -- (1.79974, 2.8898) -- (1.82217,    2.86639) -- (1.8446, 2.84049) -- (1.86703, 2.81203) -- (1.88946,    2.78091) -- (1.91189, 2.74706) -- (1.93432, 2.71037) -- (1.95675,    2.67075) -- (1.97918, 2.62807) -- (2.00161, 2.58224) -- (2.02404,    2.53313) -- (2.04647, 2.48063) -- (2.09133, 2.36493) -- (2.11376,    2.3015) -- (2.13619, 2.23416) -- (2.18105, 2.08729) -- (2.20348,    2.00752) -- (2.22591, 1.92338) -- (2.27077, 1.74153) -- (2.2932,    1.64365) -- (2.31563, 1.54104) -- (2.36049, 1.32144) -- (2.45021,    0.824691) -- (2.47182, 0.693934) -- (2.49343, 0.559092) -- (2.53664,    0.277915) -- (2.62306, -0.322681) -- (2.64466, -0.478637) -- (2.66627,  -0.635979) -- (2.70948, -0.952535) -- (2.7959, -1.57455) -- (2.81751,  -1.72339) -- (2.83911, -1.86795) -- (2.86072, -2.00737) -- (2.88233,  -2.1408) -- (2.90393, -2.2674) -- (2.92554, -2.38633) -- (2.94714,  -2.49677) -- (2.96875, -2.59795) -- (2.99035, -2.68912) -- (3.01196,  -2.76963) -- (3.03356, -2.83884) -- (3.05517, -2.89623) -- (3.07678,  -2.94136) -- (3.09838, -2.9385) -- (3.11999, -2.99345) -- (3.14159, -3) ;
    \draw [latex-latex,line width=1] (0,-3.8) -- (0,3.8);     \draw [latex-latex,line width=1] (-4,0) -- (4,0);
\draw[dashed,thin] (-3.14,-3) -- (-3.14,0); \draw[dashed,thin] (3.14,-3) -- (3.14,0);
\draw (-3.14,0.2) node {} node {\footnotesize $-\pi$}; \draw (3.14,0.2) node {} node {\footnotesize $\pi$};
    \draw (3.8,-0.3) node {} node {$\phi$};     \draw (-0.3,3.5) node {} node {$\mathcal{P}$};   \end{tikzpicture} \hspace{0.5cm} \begin{tikzpicture}[scale=1.0]   \draw [line width=0.5]  (-3.14159, -3.37275) -- (-3.12061, -3.39557) -- (-3.09963, -3.40545) --  (-3.07864, -3.40146) -- (-3.05766, -3.38277) -- (-3.03668, -3.34858) --  (-3.01569, -3.29825) -- (-2.99471, -3.23122) -- (-2.97373, -3.1471) --  (-2.95274, -3.04564) -- (-2.93176, -2.92679) -- (-2.91078, -2.79066) --  (-2.88979, -2.63758) -- (-2.86881, -2.46806) -- (-2.84783, -2.28285) --  (-2.80586, -1.86927) -- (-2.78488, -1.64335) -- (-2.76389, -1.40663) --  (-2.72193, -0.907463) -- (-2.63799, 0.140883) -- (-2.61701,    0.401821) -- (-2.59603, 0.658365) -- (-2.55406, 1.15097) -- (-2.53308,    1.38368) -- (-2.51209, 1.60535) -- (-2.47013, 2.01062) -- (-2.44738,    2.20684) -- (-2.42463, 2.38535) -- (-2.40188, 2.54556) -- (-2.37913,    2.68712) -- (-2.35639, 2.80993) -- (-2.33364, 2.91414) -- (-2.31089,    3.0001) -- (-2.28814, 3.06836) -- (-2.26539, 3.11963) -- (-2.24264,    3.15474) -- (-2.2199, 3.17467) -- (-2.19715, 3.18044) -- (-2.1744,    3.17315) -- (-2.15165, 3.15394) -- (-2.1289, 3.12396) -- (-2.10615,    3.08435) -- (-2.08341, 3.03624) -- (-2.06066, 2.98072) -- (-2.03791,    2.91883) -- (-2.01516, 2.85157) -- (-1.99241, 2.77987) -- (-1.96966,    2.70459) -- (-1.92417, 2.54646) -- (-1.90142, 2.46498) -- (-1.87867,    2.38271) -- (-1.83317, 2.21783) -- (-1.74218, 1.89751) -- (-1.72094,    1.82604) -- (-1.6997, 1.75613) -- (-1.65722, 1.6214) -- (-1.63598,    1.55668) -- (-1.61474, 1.49378) -- (-1.57225, 1.37341) -- (-1.55101,    1.31592) -- (-1.52977, 1.26017) -- (-1.48729, 1.15372) -- (-1.40233,    0.959041) -- (-1.38109, 0.913693) -- (-1.35985, 0.869491) -- (-1.31737,    0.784123) -- (-1.2324, 0.622623) -- (-1.06248, 0.311928) -- (-1.04165,    0.272811) -- (-1.02083, 0.233093) -- (-0.979179, 0.151409) -- (-0.958355,    0.109218) -- (-0.937531,    0.065978) -- (-0.895883, -0.0241019) -- (-0.875058, -0.0711666) --  (-0.854234, -0.11973) -- (-0.812586, -0.221794) -- (-0.729289,  -0.448637) -- (-0.708465, -0.51066) -- (-0.687641, -0.574987) --  (-0.645993, -0.710862) -- (-0.625169, -0.782538) -- (-0.604345,  -0.856775) -- (-0.562696, -1.01305) -- (-0.541872, -1.09511) -- (-0.521048,  -1.17977) -- (-0.4794, -1.3567) -- (-0.396103, -1.73811) -- (-0.373514,  -1.84679) -- (-0.350925, -1.95711) -- (-0.305747, -2.18093) -- (-0.283158,  -2.2934) -- (-0.260568, -2.40543) -- (-0.21539, -2.62526) -- (-0.192801,  -2.73143) -- (-0.170212, -2.83387) -- (-0.147623, -2.93157) -- (-0.125034,  -3.02346) -- (-0.102444, -3.1084) -- (-0.0798553, -3.18518) -- (-0.0572661,  -3.25257) -- (-0.034677, -3.30931) -- (-0.0120878, -3.35412) -- (0.0105013,  -3.38573) -- (0.0330905, -3.40291) -- (0.0556796, -3.40447) -- (0.0782687,  -3.38931) -- (0.100858, -3.35644) -- (0.123447, -3.30499) -- (0.146036,  -3.23426) -- (0.168625, -3.14373) -- (0.191214, -3.03312) -- (0.213804,  -2.90235) -- (0.236393, -2.75162) -- (0.258982, -2.58138) -- (0.281571,  -2.39237) -- (0.30416, -2.18558) -- (0.326749, -1.96232) -- (0.347831,  -1.74043) -- (0.368913, -1.50698) -- (0.411076, -1.01193) -- (0.495402,    0.0381647) -- (0.516484, 0.301518) -- (0.537566, 0.561182) -- (0.579729,    1.06189) -- (0.600811, 1.29943) -- (0.621892, 1.52635) -- (0.664056,    1.94302) -- (0.685137, 2.13062) -- (0.706219, 2.30329) -- (0.727301,    2.46045) -- (0.748382, 2.60173) -- (0.769464, 2.72693) -- (0.790546,    2.83604) -- (0.811627, 2.92923) -- (0.832709, 3.00682) -- (0.853791,    3.06924) -- (0.874872, 3.11708) -- (0.895954, 3.15101) -- (0.917036,    3.17177) -- (0.938117, 3.18019) -- (0.959199, 3.17713) -- (0.98028,    3.16349) -- (1.00136, 3.14017) -- (1.02421, 3.10504) -- (1.04706,    3.06078) -- (1.0699, 3.00851) -- (1.09275, 2.94931) -- (1.1156,    2.88422) -- (1.13844, 2.81421) -- (1.18414, 2.66301) -- (1.20698,    2.58344) -- (1.22983, 2.50217) -- (1.27552, 2.337) -- (1.36691,    2.00962) -- (1.38976, 1.93049) -- (1.4126, 1.85295) -- (1.4583,    1.70329) -- (1.48114, 1.63141) -- (1.50399, 1.56158) -- (1.54968,    1.42821) -- (1.57253, 1.36467) -- (1.59538, 1.30318) -- (1.64107,    1.18618) -- (1.73245, 0.973837) -- (1.75488, 0.925577) -- (1.77731,    0.878623) -- (1.82217, 0.788164) -- (1.91189, 0.617628) -- (1.93432,    0.576416) -- (1.95675, 0.535512) -- (2.00161, 0.454096) -- (2.09133,    0.289045) -- (2.11376, 0.246519) -- (2.13619, 0.203212) -- (2.18105,    0.113689) -- (2.20348, 0.0671924) -- (2.22591,    0.0193508) -- (2.27077, -0.0809295) -- (2.2932, -0.133648) -- (2.31563,  -0.188269) -- (2.36049, -0.303757) -- (2.45021, -0.563269) -- (2.47182,  -0.632081) -- (2.49343, -0.703523) -- (2.53664, -0.854557) -- (2.62306,  -1.19017) -- (2.64466, -1.28102) -- (2.66627, -1.37454) -- (2.70948,  -1.56912) -- (2.7959, -1.98287) -- (2.81751, -2.08974) -- (2.83911,  -2.19721) -- (2.88233, -2.41185) -- (2.90393, -2.51785) -- (2.92554,  -2.62208) -- (2.96875, -2.82215) -- (2.99035, -2.91629) -- (3.01196,  -3.00527) -- (3.03356, -3.08811) -- (3.05517, -3.16376) -- (3.07678,  -3.23116) -- (3.09838, -3.2892) -- (3.11999, -3.33677) -- (3.14159,  -3.37275)
;
    \draw [latex-latex,line width=1] (0,-3.8) -- (0,3.8);     \draw [latex-latex,line width=1] (-4,0) -- (4,0);
\draw[dashed,thin] (-3.14,-3.37) -- (-3.14,0); \draw[dashed,thin] (3.14,-3.37) -- (3.14,0);
\draw (-3.14,0.2) node {} node {\footnotesize $-\pi$}; \draw (3.14,0.2) node {} node {\footnotesize $\pi$};
    \draw (3.8,-0.3) node {} node {$\phi$};     \draw (-0.3,3.5) node {} node {$\mathcal{J}$};   \end{tikzpicture}
} \caption{Profiles of ${\cal P}$ (cnoidal wave in \eqref{eq:Pcnoidal}) and ${\cal J}$ in \eqref{eq:J cnoidal} for a fixed generic value of $A$, $\alpha$, $m$ and $t$.}
\label{Fig2}
\end{center}
\end{figure}

The profiles of $\mathcal{P}$ and $\mathcal{J}$ are sketched in
figure \ref{Fig2}.

As an ending remark of this section, it is worth pointing out that
in the special case of $c_{\mathcal{J}}=a=0$, the field equation
for $\mathcal{J}$ becomes devoid of a source. The particular solutions,
given by \eqref{eq:SolJamig} for the real line, and by \eqref{eq:J cnoidal}
in the case of $S^{1}$, in this case turn out to be described by
odd analytic functions that maintain its shape as they propagate with
the same velocity as their corresponding KdV solitons described by
$\mathcal{P}$ in \eqref{eq:Psoliton} and \eqref{eq:Pcnoidal}, respectively.
Note that, according to eq. \eqref{eq:EnergySolutions}, the total
energy of this sort of soliton-antisoliton bound states for $\mathcal{J}$
vanishes, and the conserved charges $\tilde{H}^{\left(n\right)}$
in \eqref{eq:Htilde} also do. Nevertheless, the conserved charges
$H_{\text{KdV}}^{\left(n\right)}$ remain being nontrivial.

\section{Geometrization of the hierarchy: the dynamics of locally flat spacetimes
in 3D\label{sec:Geometrization-of-the}}

In this section, we show that the entire structure of the class of
integrable systems with BMS$_{3}$ Poisson structure described above
can be fully geometrized, in the sense that the dynamics turns out
to be equivalently understood through the evolution of spacelike surfaces
embedded in locally flat spacetimes in three dimensions.

For the sake of simplicity, here we focus in the case of BMS$_{3}$
Poisson structures with $c_{\mathcal{J}}=0$. Thus, following the
lines of \cite{Perez:2016vqo}, it is possible to unveil a deep link
between the class of integrable systems aforementioned and General
Relativity in three spacetime dimensions. Concretely, here we show
that the Einstein-Hilbert action without cosmological constant in
3D can be endowed with an appropriate set of boundary conditions,
so that in the reduced phase space, the Einstein equations in vacuum,
which imply the vanishing of the Riemann tensor, precisely reduce
to the ones of the dynamical systems with BMS$_{3}$ Poisson structure
in \eqref{eq:EQM}. As a consequence, it is possible to establish
a one-to-one map between any solution of this kind of integrable systems
and certain specific locally flat metric in three spacetime dimensions.
Furthermore, the symmetries of the integrable systems can be seen
to naturally emerge from diffeomorphisms that preserve the asymptotic
form of the spacetime metric. Hence, and remarkably, the symmetries
manifestly become Noetherian in our geometric framework. Therefore,
the infinite set of conserved charges for the integrable system is
transparently recovered from the corresponding surface integrals in
the canonical approach. This can be seen as follows.

The Einstein-Hilbert action in three spacetime dimensions
\begin{equation}
I=\frac{1}{16\pi G}\int d^{3}x\sqrt{-g}R\,,\label{eq:EInstein-Hilbert}
\end{equation}
can be equivalently expressed as a Chern-Simons action for the $isl(2,\mathbb{R})$
algebra \cite{Achucarro:1987vz}, \cite{Witten:1988hc}. Thus, up
to boundary terms, the action \eqref{eq:EInstein-Hilbert} can be
written as
\begin{equation}
I=\frac{1}{2}\int d^{3}x\left\langle AdA+\frac{2}{3}A^{3}\right\rangle \,,\label{eq:ICS}
\end{equation}
where \textbf{$\left\langle \cdots\right\rangle $} stands for the
invariant bilinear form defined in eq. \eqref{eq:bilinear metric}
with $c_{\mathcal{J}}=0$ and $c_{\mathcal{P}}=1/\left(8\pi G\right)$.
The components of the $isl(2,\mathbb{R})$-valued gauge field are
then identified with the dualized spin connection and the dreibein
according to 

\begin{equation}
A=\omega^{a}J_{a}+e^{a}P_{a}\,.\label{eq:vielbein}
\end{equation}

In order to describe the asymptotic structure of the fields, as explained
in \cite{Coussaert:1995zp}, \cite{Henneaux:2013dra}, \cite{Bunster:2014mua}
it is useful to choose the gauge so that the connection reads
\begin{equation}
A=b^{-1}ab+b^{-1}db\,,\label{eq:Agrande}
\end{equation}
where the radial dependence is completely captured by the group element
$b=b\left(r\right)$, which as shown in \cite{Matulich:2014hea} can
be conveniently chosen as $b=e^{rP_{2}}$. One of the advantages of
this gauge choice is that the remaining analysis can be performed
in terms of the auxiliary connection 
\begin{equation}
a=a_{t}dt+a_{\phi}d\phi\;,\label{eq:achico}
\end{equation}
that only depends on $t$, $\phi$. Here we propose that the asymptotic
form of the auxiliary connection for the gravitational field in \eqref{eq:achico}
is precisely given by the two-dimensional locally flat gauge field
that describes the field equations of the dynamical system with BMS$_{3}$
Poisson structure in section \ref{sec:Drinfeld-Sokolov-formulation}.
Their components $a_{\phi}$ and $a_{t}$ are then described by eqs.
\eqref{eq:a-phi} and \eqref{eq:at}, respectively. Therefore, from
\eqref{eq:ICS}, the field equations imply that the connection $A$
is flat ($F=dA+A^{2}=0$), which by virtue of \eqref{eq:vielbein}
amounts to deal with three-dimensional manifolds with vanishing curvature
and torsion; whereas eq. \eqref{eq:Agrande} means that the field
strength of the auxiliary gauge field also vanishes as in \eqref{eq:fchico}.
Hence, for our boundary conditions, the Einstein equations in vacuum
precisely reduce to the ones of a dynamical system with BMS$_{3}$
Poisson structure in \eqref{eq:Dynamical-system}.

Besides, the asymptotic symmetries, being defined as the diffeomorphisms
that preserve the asymptotic form of the spacetime metric, turn out
to be equivalent to the set of gauge transformations $\delta A=d\tilde{\lambda}+[A,\tilde{\lambda}]$
that maintain the asymptotic form of the gauge field in \eqref{eq:Agrande}.
For our boundary conditions, one then finds that the asymptotic symmetries
are spanned by a Lie-algebra-valued parameter of the form $\tilde{\lambda}=b^{-1}\lambda b$,
where $\lambda=\Lambda\left(\varepsilon_{\mathcal{J}},\varepsilon_{\mathcal{P}}\right)$
is exactly given as in eq. \eqref{eq:lambda-chico}, including the
consistency condition for the parameters $\varepsilon_{\mathcal{J}}$,
$\varepsilon_{\mathcal{P}}$ in \eqref{eq:epsilon-punto}. Furthermore,
the transformation law of the dynamical fields ${\cal J}$, ${\cal P}$
precisely agrees with the transformations in eq. \eqref{eq:delta-P}.

Since the asymptotic symmetries are Noetherian, the global conserved
charges can be readily obtained using the canonical approach \cite{Regge:1974zd}.
Indeed, their variation is explicitly given by surface integrals defined
at the boundary of the spatial section,\footnote{It is worth noting that the boundary can be located at any fixed value
of the radial coordinate, and hence, not necessarily at null infinity.} which read
\begin{equation}
\delta Q[\tilde{\lambda}]=-\int d\phi\left\langle \tilde{\lambda}\delta A_{\phi}\right\rangle =-\int d\phi\left\langle \lambda\delta a_{\phi}\right\rangle =-\int d\phi\left(\varepsilon_{\mathcal{J}}\delta\mathcal{J}+\varepsilon_{\mathcal{P}}\delta\mathcal{P}\right)\;,\label{eq:deltaqg}
\end{equation}
and precisely coincide with the variation of the conserved charges
introduced in eq. \eqref{eq:canoncalQ}, in the context of integrable
systems. 

In particular, if the Lagrange multipliers $\mu_{\mathcal{J}}$ and
$\mu_{\mathcal{P}}$ are kept fixed at the boundary according to \eqref{eq:musk},
so that they correspond to the variation of the functional $H^{\left(k\right)}$
defined in \eqref{eq:Hk}, the Einstein equations reduce to the ones
of the hierarchy of integrable systems discussed in section \ref{subsec:The-hierarchyk>1}.
Therefore, in this case, the variation of the surface integrals in
\eqref{eq:deltaqg} integrates precisely as in \eqref{eq:Q-1}. Note
that the energy of a gravitational configuration that fulfills these
boundary conditions is then given by the Hamiltonian of the corresponding
integrable system as in \eqref{eq:Energy}, i.e., $E=Q[\partial_{t}]=H^{\left(k\right)}$. 

In the simplest case of $k=0$, described in section \ref{subsec:Warming-up-with},
we recover the set of boundary conditions proposed in \cite{Matulich:2014hea}\footnote{Strictly speaking, we are dealing with the boundary conditions in
\cite{Matulich:2014hea} provided that the higher spin fields and
their corresponding chemical potentials are turned-off.} (see also \cite{Gary:2014ppa}) which contain the boundary conditions
in \cite{Barnich:2006av} for a particular choice of Lagrange multipliers
at the boundary. Note that in this case, the BMS$_{3}$ algebra is
realized as the asymptotic symmetry algebra. For the remaining cases
($k\geq1$), the new class of boundary conditions is such that the
asymptotic symmetry algebra is infinite-dimensional, abelian and devoid
of central charges, which is equivalent to the fact that the conserved
charges of the hierarchy are in involution (see eq. \eqref{eq:involution}).

It is worth pointing out that, in the metric formalism, our particular
choices for the Lagrange multipliers $\mu_{\mathcal{J}}$ and $\mu_{\mathcal{P}}$
correspond to fixing the lapse and shift functions in the ADM decomposition
of the metric, as suitable local functionals of the dynamical variables
at the boundary. Furthermore, and remarkably, the class of locally
flat spacetimes described by this new set of asymptotic conditions
inherits the anisotropic Lifshitz scaling of the corresponding integrable
systems by virtue of the boundary conditions. This effect then becomes
the flat analogue of the one found in \cite{Perez:2016vqo} for the
case of locally AdS$_{3}$ spacetimes, where isotropy is recovered
in the particular case of the Brown-Henneaux boundary conditions \cite{Brown:1986nw}.

It is also amusing to verify that trivial solutions of the class of
integrable systems described above, like the ones just given by configurations
with ${\cal J}$ and ${\cal P}$ constants, actually correspond to
the geometries of the flat cosmological spacetimes in \cite{Cornalba:2002fi}.
Hence, from a gravitational point of view, these configurations become
certainly non-trivial, not only in a geometric sense, but also in
the physical one since they carry Hawking temperature and entropy
associated to the cosmological horizon. It would then be worth to
explore additional simple configurations for the integrable systems
that could naturally become non-trivial in the gravitational framework
and vice versa. 

The precise details of the asymptotic structure in the metric formulation,
including the study of some of the physical properties of the flat
cosmological spacetimes once they are embedded within the new set
of boundary conditions presented in this section will be discussed
in a forthcoming work. It can also be shown that the geometrization
of the class of integrable systems discussed here, but in the generic
case with $c_{\mathcal{J}}\neq0$, can be performed through a suitable
extension of the analysis in \cite{Barnich:2014cwa}, \cite{Fuentealba:2015wza}
once suitable parity odd terms in the action are included (see also
\cite{Giacomini:2006dr}). 

\section{Extensions of our results\label{sec:Extensions}}

As pointed out in the introduction, the BMS$_{3}$ algebra admits
some extensions which, according to our results, might be expected
to be linked to new classes of integrable systems. In particular,
an interesting nonlinear extension of BMS$_{3}$ that includes additional
generators of spin $s>2$ was found in \cite{Gonzalez:2013oaa} (in
full agreement with the algebra simultaneously found in \cite{Afshar:2013vka}
for $s=3$). This kind of extensions can be regarded as ``flat $W$-algebras'',
since they can be recovered from a suitable Inönü-Wigner contraction
of two copies of certain classical $W$-algebras (for a review about
$W$-algebras, see e.g. \cite{Bouwknegt:1992wg}). Noteworthy, preliminary
results \cite{W-Flat-IntegrableSystems} show that new hierarchies
of integrable systems whose Poisson structures correspond to flat
$W$-algebras indeed exist. In fact, this family of integrable systems
turns out to be bi-Hamiltonian, and furthermore, following the lines
of section \ref{sec:Geometrization-of-the}, they can also be geometrized
in terms of higher spin gravity without cosmological constant in three
spacetime dimensions endowed with a suitable set of boundary conditions.
Therefore, the symmetries of this novel class of integrable systems
can be seen as combinations of diffeomorphisms and higher spin gauge
transformations that preserve the asymptotic form of the three-dimensional
configurations. As a consequence, in the three-dimensional geometric
setup, the infinite set of conserved charges of the integrable systems
emerge as the canonical generators that correspond to the asymptotic
symmetries, being described by suitable surface integrals at the boundary,
which turn out to be in involution.

Further interesting links between certain well-known classes of integrable
systems and higher spin gravity on AdS$_{3}$ have been explored in
\cite{Compere:2013gja}, \cite{Gutperle:2014aja}, \cite{Beccaria:2015iwa},
\cite{Perez:2016vqo}, \cite{Gutperle:2017ewo}.

\acknowledgments We thank Rafael Benguria, Marcela Cárdenas, Hernán
A. González, Daniel Grumiller, Dmitry Melnikov and Fábio Novaes for
useful discussions and comments. We are indebted to Cédric Troessaert
for helpful remarks and collaboration in an early stage of this work.
The work of P.R. was partially funded by the PhD grant CONICYT-PCHA/Doctorado
Nacional/2016-21161262. The work of J.M. was supported by the ERC
Advanced Grant ``High-Spin-Grav'', by FNRS-Belgium (convention FRFC
PDR T.1025.14 and convention IISN 4.4503.15). This research has been
partially supported by Fondecyt grants Nº 3150448, 3170772, 11130083,
11130260, 11130262, 1161311, 1171162. The Centro de Estudios Científicos
(CECs) is funded by the Chilean Government through the Centers of
Excellence Base Financing Program of Conicyt.

\appendix

\section{List of conserved quantities and polynomials \label{sec:Lists} }

Here we provide an explicit list of the first six conserved charges
$H_{\text{KdV}}^{\left(n\right)}$ and $\tilde{H}^{\left(n\right)}$,
and their associated polynomials $R^{(n)}$ and $T^{(n)}$. 

The conserved quantities $H_{\text{KdV}}^{\left(n\right)}$ correspond
to the ones of the KdV equation in \eqref{eq:EoMJP1}, which in our
conventions, generically depend on the central extension $c_{\mathcal{P}}$.
Therefore, they read
\begin{eqnarray*}
H_{\text{KdV}}^{\left(0\right)} & = & \int d\phi\mathcal{P}\,,\\
H_{\text{KdV}}^{\left(1\right)} & = & \int d\phi\frac{\mathcal{P}^{2}}{2}\,,\\
H_{\text{KdV}}^{\left(2\right)} & = & \int d\phi\left[\frac{1}{2}c_{\mathcal{P}}\mathcal{P}'^{2}+\frac{1}{2}\mathcal{P}^{3}\right]\,,\\
H_{\text{KdV}}^{\left(3\right)} & = & \int d\phi\left[\frac{1}{2}c_{\mathcal{P}}^{2}\mathcal{P}''^{2}+\frac{5}{2}c_{\mathcal{P}}\mathcal{P}\mathcal{P}'^{2}+\frac{5}{8}\mathcal{P}^{4}\right]\,,\\
H_{\text{KdV}}^{\left(4\right)} & = & \int d\phi\left[\frac{1}{2}c_{\mathcal{P}}^{3}(\mathcal{P}^{(3)}){}^{2}+\frac{7}{2}c_{\mathcal{P}}^{2}\mathcal{P}\mathcal{P}''^{2}+\frac{35}{4}c_{\mathcal{P}}\mathcal{P}^{2}\mathcal{P}'^{2}+\frac{7}{8}\mathcal{P}^{5}\right]\,,\\
H_{\text{KdV}}^{\left(5\right)} & = & \int d\phi\Big[\frac{1}{2}c_{\mathcal{P}}^{4}(\mathcal{P}^{(4)}){}^{2}+\frac{9}{2}c_{\mathcal{P}}^{3}(\mathcal{P}^{(3)}){}^{2}\mathcal{P}+\frac{63}{4}c_{\mathcal{P}}^{2}\mathcal{P}^{2}\mathcal{P}''^{2}-5c_{\mathcal{P}}^{3}\mathcal{P}''^{3}\\
 &  & +\frac{105}{4}c_{\mathcal{P}}\mathcal{P}^{3}\mathcal{P}'^{2}-\frac{35}{8}c_{\mathcal{P}}^{2}\mathcal{P}'^{4}+\frac{21}{16}\mathcal{P}^{6}\Big]\,.
\end{eqnarray*}
The remaining conserved charges $\tilde{H}^{(n)}$, can then be readily
obtained from \eqref{eq:Htilde}, which are given by
\begin{eqnarray*}
\tilde{H}^{(0)} & = & \int d\phi\mathcal{J}\,,\\
\tilde{H}^{(1)} & = & \int d\phi\mathcal{J}\mathcal{P}\,,\\
\tilde{H}^{(2)} & = & \int d\phi\Big[\mathcal{J}\Big(\frac{3}{2}\mathcal{P}^{2}-c_{\mathcal{P}}\mathcal{P}''\Big)+\frac{1}{2}c_{\mathcal{J}}\mathcal{P}'^{2}\Big]\,,\\
\tilde{H}^{(3)} & = & \int d\phi\Big[\mathcal{J}\Big(c_{\mathcal{P}}^{2}\mathcal{P}^{(4)}-5c_{\mathcal{P}}\mathcal{P}\mathcal{P}''-\frac{5}{2}c_{\mathcal{P}}\mathcal{P}'^{2}+\frac{5}{2}\mathcal{P}^{3}\Big)+c_{\mathcal{J}}\Big(c_{\mathcal{P}}\mathcal{P}''^{2}+\frac{5}{2}\mathcal{P}\mathcal{P}'^{2}\Big)\Big]\,,\\
\tilde{H}^{(4)} & = & \int d\phi\Big[\mathcal{J}\Big(-c_{\mathcal{P}}^{3}\mathcal{P}^{(6)}+7c_{\mathcal{P}}^{2}\mathcal{P}^{(4)}\mathcal{P}-\frac{35}{2}c_{\mathcal{P}}\mathcal{P}^{2}\mathcal{P}''+\frac{21}{2}c_{\mathcal{P}}^{2}\mathcal{P}''^{2}-\frac{35}{2}c_{\mathcal{P}}\mathcal{P}\mathcal{P}'^{2}\\
 &  & +14c_{\mathcal{P}}^{2}\mathcal{P}^{(3)}\mathcal{P}'+\frac{35}{8}\mathcal{P}^{4}\Big)+c_{\mathcal{J}}\Big(\frac{3}{2}c_{\mathcal{P}}^{2}(\mathcal{P}^{(3)})^{2}+7c_{\mathcal{P}}\mathcal{P}\mathcal{P}''^{2}+\frac{35}{4}\mathcal{P}^{2}\mathcal{P}'^{2}\Big)\Big]\,,\\
\tilde{H}^{(5)} & = & \int d\phi\Big[\mathcal{J}\Big(c_{\mathcal{P}}^{4}\mathcal{P}^{(8)}-9c_{\mathcal{P}}^{3}\mathcal{P}^{(6)}\mathcal{P}-\frac{69}{2}c_{\mathcal{P}}^{3}(\mathcal{P}^{(3)})^{2}+\frac{189}{2}c_{\mathcal{P}}^{2}\mathcal{P}\mathcal{P}''^{2}\\
 &  & -27c_{\mathcal{P}}^{3}\mathcal{P}^{(5)}\mathcal{P}'-57c_{\mathcal{P}}^{3}\mathcal{P}^{(4)}\mathcal{P}''+\frac{63}{2}c_{\mathcal{P}}^{2}\mathcal{P}^{(4)}\mathcal{P}^{2}-\frac{105}{2}c_{\mathcal{P}}\mathcal{P}^{3}\mathcal{P}''\\
 &  & +126c_{\mathcal{P}}^{2}\mathcal{P}^{(3)}\mathcal{P}\mathcal{P}'+\frac{231}{2}c_{\mathcal{P}}^{2}\mathcal{P}'^{2}\mathcal{P}''-\frac{315}{4}c_{\mathcal{P}}\mathcal{P}^{2}\mathcal{P}'^{2}+\frac{63\mathcal{P}^{5}}{8}\Big)\\
 &  & +c_{\mathcal{J}}\Big(2c_{\mathcal{P}}^{3}(\mathcal{P}^{(4)})^{2}+\frac{27}{2}c_{\mathcal{P}}^{2}\mathcal{P}(\mathcal{P}^{(3)})^{2}-15c_{\mathcal{P}}^{2}\mathcal{P}''^{3}+\frac{63}{2}c_{\mathcal{P}}\mathcal{P}^{2}\mathcal{P}''^{2}\\
 &  & -\frac{35}{4}c_{\mathcal{P}}\mathcal{P}'^{4}+\frac{105}{4}\mathcal{P}^{3}\mathcal{P}'^{2}\Big)\Big]\,.
\end{eqnarray*}

Hence, according to \eqref{eq:Rn-1}, the Gelfand-Dikii polynomials
read
\begin{eqnarray*}
R^{(0)} & = & 1\,,\quad R^{(1)}=\mathcal{P}\,,\quad R^{(2)}=-c_{\mathcal{P}}\mathcal{P}''+\frac{3}{2}\mathcal{P}^{2}\,,\\
R^{(3)} & = & c_{\mathcal{P}}^{2}\mathcal{P}^{(4)}-5c_{\mathcal{P}}\mathcal{P}\mathcal{P}''-\frac{5}{2}c_{\mathcal{P}}\mathcal{P}'{}^{2}+\frac{5}{2}\mathcal{P}{}^{3}\,,\\
R^{(4)} & = & -c_{\mathcal{P}}^{3}\mathcal{P}^{(6)}+7c_{\mathcal{P}}^{2}\mathcal{P}^{(4)}\mathcal{P}-\frac{35}{2}c_{\mathcal{P}}\mathcal{P}{}^{2}\mathcal{P}''+\frac{21}{2}c_{\mathcal{P}}^{2}\mathcal{P}''{}^{2}-\frac{35}{2}c_{\mathcal{P}}\mathcal{P}\mathcal{P}'{}^{2}\\
 &  & +14c_{\mathcal{P}}^{2}\mathcal{P}^{(3)}\mathcal{P}'+\frac{35}{8}\mathcal{P}{}^{4}\,,\\
R^{(5)} & = & c_{\mathcal{P}}^{4}\mathcal{P}^{(8)}-9c_{\mathcal{P}}^{3}\mathcal{P}^{(6)}\mathcal{P}+\frac{63}{2}c_{\mathcal{P}}^{2}\mathcal{P}^{(4)}\mathcal{P}{}^{2}-\frac{69}{2}c_{\mathcal{P}}^{3}(\mathcal{P}^{(3)}){}^{2}-\frac{105}{2}c_{\mathcal{P}}\mathcal{P}{}^{3}\mathcal{P}''\\
 &  & +\frac{189}{2}c_{\mathcal{P}}^{2}\mathcal{P}\mathcal{P}''{}^{2}-\frac{315}{4}c_{\mathcal{P}}\mathcal{P}{}^{2}\mathcal{P}'{}^{2}-27c_{\mathcal{P}}^{3}\mathcal{P}^{(5)}\mathcal{P}'-57c_{\mathcal{P}}^{3}\mathcal{P}^{(4)}\mathcal{P}''\\
 &  & +126c_{\mathcal{P}}^{2}\mathcal{P}^{(3)}\mathcal{P}\mathcal{P}'+\frac{231}{2}c_{\mathcal{P}}^{2}\mathcal{P}'{}^{2}\mathcal{P}''+\frac{63}{8}\mathcal{P}{}^{5}\,,
\end{eqnarray*}
while the polynomials $T^{(n)}$ can be obtained from \eqref{eq:Tn-1},
so that they are given by
\begin{eqnarray*}
T^{(0)} & = & 0\,,\quad T^{(1)}=\mathcal{J}\,,\quad T^{(2)}=-c_{\mathcal{P}}\mathcal{J}''-c_{\mathcal{J}}\mathcal{P}''+3\mathcal{J}\mathcal{P}\,,\\
T^{(3)} & = & c_{\mathcal{P}}^{2}\mathcal{J}^{(4)}-5c_{\mathcal{P}}\mathcal{P}\mathcal{J}''-5c_{\mathcal{P}}\mathcal{J}\mathcal{P}''-5c_{\mathcal{P}}\mathcal{J}'\mathcal{P}'+\frac{15}{2}\mathcal{J}\mathcal{P}^{2}\\
 &  & +c_{\mathcal{J}}\left(2c_{\mathcal{P}}\mathcal{P}^{(4)}-5\mathcal{P}\mathcal{P}''-\frac{5}{2}\mathcal{P}'^{2}\right)\,,\\
T^{(4)} & = & -c_{\mathcal{P}}^{3}\mathcal{J}^{(6)}+7c_{\mathcal{P}}^{2}\mathcal{J}^{(4)}\mathcal{P}+14c_{\mathcal{P}}^{2}\mathcal{J}^{(3)}\mathcal{P}'+21c_{\mathcal{P}}^{2}\mathcal{J}''\mathcal{P}''-\frac{35}{2}c_{\mathcal{P}}\mathcal{P}^{2}\mathcal{J}''+14c_{\mathcal{P}}^{2}\mathcal{P}^{(3)}\mathcal{J}'\\
 &  & +7c_{\mathcal{P}}^{2}\mathcal{J}\mathcal{P}^{(4)}-35c_{\mathcal{P}}\mathcal{P}\mathcal{J}'\mathcal{P}'-35c_{\mathcal{P}}\mathcal{J}\mathcal{P}\mathcal{P}''-\frac{35}{2}c_{\mathcal{P}}\mathcal{J}\mathcal{P}'^{2}+\frac{35}{2}\mathcal{J}\mathcal{P}^{3}-c_{\mathcal{J}}\Big(3c_{\mathcal{P}}^{2}\mathcal{P}^{(6)}\\
 &  & -14c_{\mathcal{P}}\mathcal{P}\mathcal{P}^{(4)}-21c_{\mathcal{P}}\mathcal{P}''^{2}+\frac{35}{2}\mathcal{P}^{2}\mathcal{P}''+\frac{35}{2}\mathcal{P}\mathcal{P}'^{2}-28c_{\mathcal{P}}\mathcal{P}^{(3)}\mathcal{P}'\Big)\,,\\
T^{(5)} & = & c_{\mathcal{P}}^{4}\mathcal{J}^{(8)}-9c_{\mathcal{P}}^{3}\mathcal{J}^{(6)}\mathcal{P}-27c_{\mathcal{P}}^{3}\mathcal{J}^{(5)}\mathcal{P}'-57c_{\mathcal{P}}^{3}\mathcal{J}^{(4)}\mathcal{P}''+\frac{63}{2}c_{\mathcal{P}}^{2}\mathcal{J}^{(4)}\mathcal{P}^{2}-69c_{\mathcal{P}}^{3}\mathcal{J}^{(3)}\mathcal{P}^{(3)}\\
 &  & +126c_{\mathcal{P}}^{2}\mathcal{J}^{(3)}\mathcal{P}\mathcal{P}'-57c_{\mathcal{P}}^{3}\mathcal{P}^{(4)}\mathcal{J}''+189c_{\mathcal{P}}^{2}\mathcal{P}\mathcal{J}''\mathcal{P}''+\frac{231}{2}c_{\mathcal{P}}^{2}\mathcal{J}''\mathcal{P}'^{2}-\frac{105}{2}c_{\mathcal{P}}\mathcal{P}^{3}\mathcal{J}''\\
 &  & -27c_{\mathcal{P}}^{3}\mathcal{P}^{(5)}\mathcal{J}'+126c_{\mathcal{P}}^{2}\mathcal{P}\mathcal{P}^{(3)}\mathcal{J}'-\frac{315}{2}c_{\mathcal{P}}\mathcal{P}^{2}\mathcal{J}'\mathcal{P}'+231c_{\mathcal{P}}^{2}\mathcal{J}'\mathcal{P}'\mathcal{P}''-9c_{\mathcal{P}}^{3}\mathcal{J}\mathcal{P}^{(6)}\\
 &  & +63c_{\mathcal{P}}^{2}\mathcal{J}\mathcal{P}\mathcal{P}^{(4)}+\frac{189}{2}c_{\mathcal{P}}^{2}\mathcal{J}\mathcal{P}''^{2}-\frac{315}{2}c_{\mathcal{P}}\mathcal{J}\mathcal{P}^{2}\mathcal{P}''-\frac{315}{2}c_{\mathcal{P}}\mathcal{J}\mathcal{P}\mathcal{P}'^{2}+126c_{\mathcal{P}}^{2}\mathcal{J}\mathcal{P}^{(3)}\mathcal{P}'\\
 &  & +\frac{315}{8}\mathcal{J}\mathcal{P}^{4}+c_{\mathcal{J}}\Big(4c_{\mathcal{P}}^{3}\mathcal{P}^{(8)}-27c_{\mathcal{P}}^{2}\mathcal{P}\mathcal{P}^{(6)}+63c_{\mathcal{P}}\mathcal{P}^{2}\mathcal{P}^{(4)}-\frac{207}{2}c_{\mathcal{P}}^{2}(\mathcal{P}^{(3)}){}^{2}+189c_{\mathcal{P}}\mathcal{P}\mathcal{P}''^{2}\\
 &  & -81c_{\mathcal{P}}^{2}\mathcal{P}^{(5)}\mathcal{P}'-171c_{\mathcal{P}}^{2}\mathcal{P}^{(4)}\mathcal{P}''+252c_{\mathcal{P}}\mathcal{P}\mathcal{P}^{(3)}\mathcal{P}'+231c_{\mathcal{P}}\mathcal{P}'^{2}\mathcal{P}''-\frac{105}{2}\mathcal{P}^{3}\mathcal{P}''-\frac{315}{4}\mathcal{P}^{2}\mathcal{P}'^{2}\Big)\,.\\
\end{eqnarray*}

\section{Involution of the conserved quantities\label{sec:Involution-of-the}}

In order to prove that our set of conserved charges, $H^{(m)}=\left(H_{\text{KdV}}^{\left(n\right)};\tilde{H}^{\left(n\right)}\right)$,
is abelian in both Poisson brackets, one can follow the lines of the
proof of the same statement in the case of the pure KdV equation that
can be found in standard textbooks (see e.g., \cite{das1989integrable,olver2000applications}).

Without loss of generality, let us assume that $m>n$. Thus, the Poisson
bracket associated to the BMS$_{3}$ operator $\mathcal{D}^{(2)}$
of two conserved charges, which reads
\begin{equation}
\lbrace H^{(m)},H^{(n)}\rbrace_{(2)}=\int d\phi\left(\begin{array}{cc}
\frac{\delta H^{(m)}}{\delta\mathcal{J}} & \frac{\delta H^{(m)}}{\delta\mathcal{P}}\end{array}\right)\mathcal{D}^{\left(2\right)}\left(\begin{array}{c}
\frac{\delta H^{(n)}}{\delta\mathcal{J}}\\
\frac{\delta H^{(n)}}{\delta\mathcal{P}}
\end{array}\right)\,,
\end{equation}
by virtue of the recursion relation in \eqref{eq:recrelation}, can
be written in terms of the Poisson bracket associated to the ``canonical''
operator $\mathcal{D}^{(1)}$, according to
\begin{eqnarray}
\lbrace H^{(m)},H^{(n)}\rbrace_{(2)} & = & \int d\phi\left(\begin{array}{cc}
\frac{\delta H^{(m)}}{\delta\mathcal{J}} & \frac{\delta H^{(m)}}{\delta\mathcal{P}}\end{array}\right)\mathcal{D}^{\left(1\right)}\left(\begin{array}{c}
\frac{\delta H^{(n+1)}}{\delta\mathcal{J}}\\
\frac{\delta H^{(n+1)}}{\delta\mathcal{P}}
\end{array}\right)\nonumber \\
 & = & \lbrace H^{(m)},H^{(n+1)}\rbrace_{(1)}\\
 & = & -\lbrace H^{(n+1)},H^{(m)}\rbrace_{(1)}\,.\nonumber 
\end{eqnarray}
Analogously, making use of the recursion relationship again, one finds
that
\begin{equation}
\lbrace H^{(m)},H^{(n)}\rbrace_{(2)}=-\lbrace H^{(n+1)},H^{(m)}\rbrace_{(1)}=\lbrace H^{(m-1)},H^{(n+1)}\rbrace_{(2)}\,.
\end{equation}
Therefore, once the procedure is applied $m-n$ times, one obtains
\begin{equation}
\lbrace H^{(m)},H^{(n)}\rbrace_{(2)}=\lbrace H^{(n)},H^{(m)}\rbrace_{(2)}\;,
\end{equation}
 which implies that the conserved charges are involution in both Poisson
brackets, i.e., 
\begin{equation}
\lbrace H^{(m)},H^{(n)}\rbrace_{(2)}=\lbrace H^{(m)},H^{(n)}\rbrace_{(1)}=0\,.
\end{equation}

\section{Generic solution for the field equations of the hierarchy\label{sec:Proof}}

Here we show that the generic solution in \eqref{eq:Jgeneral} solves
the field equations \eqref{eq:EoMJPk} for an arbitrary value of the
label of the hierarchy $k$, provided that $\mathcal{P}$ stands for
an arbitrary generic solution for the field equations of the $k$-th
representative of the KdV hierarchy.

In sum, we want to prove that
\begin{equation}
\mathcal{J}=\sum_{j=0}^{\infty}\eta_{j}\partial_{\phi}R^{(j+1)}+c_{\mathcal{J}}\frac{\partial\mathcal{P}}{\partial c_{\mathcal{P}}}+at\dot{{\cal P}}\,,\label{eq:SolJ-1-1}
\end{equation}
is a solution of
\begin{equation}
\dot{\mathcal{J}}=\mathcal{D}^{(\mathcal{P})}T^{(k)}+\left(\mathcal{D}^{(\mathcal{J})}+a\mathcal{D}^{\left(\mathcal{P}\right)}\right)R^{\left(k\right)}\,,\label{eq:EoMJ-1}
\end{equation}
provided that $\mathcal{P}$ solves
\begin{equation}
\dot{\mathcal{P}}=\mathcal{D}^{\left(\mathcal{P}\right)}R^{\left(k\right)}\,,\label{eq:dotPappe}
\end{equation}
where the Gelfand-Dikii polynomials $R^{(k)}$, and the polynomials
$T^{(k)}$ are defined through \eqref{eq:Rn-1} and \eqref{eq:Tn-1},
with $\tilde{H}^{(k)}$ given by \eqref{eq:Htilde}, i.e.,
\begin{eqnarray}
R^{(k)} & = & \frac{\delta H_{\text{KdV}}^{\left(k\right)}\left[\mathcal{P}\right]}{\delta\mathcal{P}}\,,\label{eq:Rappen}\\
T^{(k)} & = & c_{\mathcal{J}}\frac{\partial R^{(k)}}{\partial c_{\mathcal{P}}}+\int d\varphi{\cal J}(\varphi)\frac{\delta R^{(k)}(\varphi)}{\delta\mathcal{P}}\,.\label{eq:Tappen}
\end{eqnarray}

Thus, once \eqref{eq:Tappen} is evaluated on \eqref{eq:SolJ-1-1},
it reduces to
\begin{equation}
T^{(k)}=\sum_{j=0}^{\infty}\eta_{j}\int d\varphi\partial_{\varphi}R^{(j+1)}(\varphi)\frac{\delta R^{(k)}}{\delta\mathcal{P}(\varphi)}+c_{\mathcal{J}}\frac{dR^{(k)}}{dc_{\mathcal{P}}}+at\partial_{t}R^{(k)}\,,
\end{equation}
where $\frac{dR^{(k)}}{dc_{\mathcal{P}}}$ stands for the total derivative
of $R^{(k)}$ with respect to the central charge $c_{\mathcal{P}}$,
given by
\begin{equation}
\frac{dR^{(k)}}{dc_{\mathcal{P}}}=\frac{\partial R^{(k)}}{\partial c_{\mathcal{P}}}+\int d\varphi\frac{\partial\mathcal{P}(\varphi)}{\partial c_{\mathcal{P}}}\frac{\delta R^{(k)}}{\delta\mathcal{P}(\varphi)}\,.
\end{equation}
Besides, the time derivative of \eqref{eq:SolJ-1-1} can be written
as
\begin{eqnarray}
\dot{\mathcal{J}} & = & \sum_{j=0}^{\infty}\eta_{j}\partial_{\phi}\left[\int d\varphi\frac{\delta R^{(j+1)}}{\delta\mathcal{P}(\varphi)}\dot{\mathcal{P}}(\varphi)\right]+c_{\mathcal{J}}\frac{\partial\dot{\mathcal{P}}}{\partial c_{\mathcal{P}}}+a\partial_{t}\left(t\partial_{t}P\right)\nonumber \\
 & = & a\mathcal{D}^{\left(\mathcal{P}\right)}R^{\left(k\right)}\nonumber \\
 &  & +\partial_{\phi}\left[-\sum_{j=0}^{\infty}\eta_{j}\int d\varphi\frac{\delta\partial_{\varphi}R^{(j+1)}}{\delta\mathcal{P}(\varphi)}R^{\left(k+1\right)}(\varphi)+c_{\mathcal{J}}\frac{\partial R^{\left(k+1\right)}}{\partial c_{\mathcal{P}}}+at\partial_{t}R^{(k+1)}\right]\,,\nonumber \\
\label{eq:dotJappen}
\end{eqnarray}
where we have made use of the $k$-th KdV equation in \eqref{eq:dotPappe},
as well as the recursion relation for the Gelfand-Dikii polynomials
in \eqref{eq:Gelfan-1}. Note that by virtue of \eqref{eq:Rappen},
the following identity holds
\begin{equation}
\int d\varphi\frac{\delta\partial_{\varphi}R^{(j+1)}}{\delta\mathcal{P}(\varphi)}R^{\left(k+1\right)}(\varphi)=\frac{\delta}{\delta\mathcal{P}}\int d\varphi\partial_{\varphi}R^{(j+1)}(\varphi)R^{\left(k+1\right)}(\varphi)-\int d\varphi\partial_{\varphi}R^{(j+1)}(\varphi)\frac{\delta R^{\left(k+1\right)}(\varphi)}{\delta\mathcal{P}}\,,\label{eq:Identity}
\end{equation}
where the first term in the r.h.s. of \eqref{eq:Identity} vanishes
due to the fact that the conserved charges $H_{\text{KdV}}^{(k)}$
are in involution, i.e., $\{H_{\text{KdV}}^{(k+1)},H_{\text{KdV}}^{(j+1)}\}_{(1)}=0$.
Hence, eq. \eqref{eq:Identity} implies that \eqref{eq:dotJappen}
reduces to
\begin{eqnarray}
\dot{\mathcal{J}} & = & a\mathcal{D}^{\left(\mathcal{P}\right)}R^{\left(k\right)}\nonumber \\
 &  & +\partial_{\phi}\left[\sum_{j=0}^{\infty}\eta_{j}\int d\varphi\partial_{\varphi}R^{(j+1)}(\varphi)\frac{\delta R^{\left(k+1\right)}(\varphi)}{\delta\mathcal{P}}+c_{\mathcal{J}}\frac{\partial R^{\left(k+1\right)}}{\partial c_{\mathcal{P}}}+at\partial_{t}R^{(k+1)}\right]\nonumber \\
 & = & a\mathcal{D}^{\left(\mathcal{P}\right)}R^{\left(k\right)}+\partial_{\phi}T^{(k+1)}\,.\label{eq:dotJappen2}
\end{eqnarray}
Therefore, making use of the recursion relation for the polynomials
$T^{(k)}$ in \eqref{Gelfan_Tempo-1}, one finally proves that eq.
\eqref{eq:dotJappen2} reduces to the field equation in \eqref{eq:EoMJ-1},
which implies that $\mathcal{J}$ in \eqref{eq:SolJ-1-1} is indeed
a solution.

Consequently, making $\eta_{j}=0$ in \eqref{eq:SolJ-1-1}, one concludes
that $\mathcal{J}_{p}$ in \eqref{eq:Jparticular} provides a particular
solution for the field equation \eqref{eq:EoMJ-1}.

\medskip{}

The conserved charges associated to this exact solution are then given
by $H_{\text{KdV}}^{(n)}$ and $\tilde{H}^{(n)}$ defined in \eqref{eq:Htilde}.
Note that once $\tilde{H}^{(n)}$ is evaluated on the exact solution
\eqref{eq:SolJ-1-1}, the contribution due to the homogeneous part
vanishes, because
\begin{equation}
\int d\phi{\cal J}_{h}\frac{\delta H_{\text{KdV}}^{\left(n\right)}}{\delta\mathcal{P}}=\sum_{j=0}^{\infty}\eta_{j}\int d\phi\partial_{\phi}R^{(j+1)}\frac{\delta H_{\text{KdV}}^{\left(n\right)}}{\delta\mathcal{P}}=\sum_{j=0}^{\infty}\eta_{j}\left\{ H_{\text{KdV}}^{(n)},H_{\text{KdV}}^{(j+1)}\right\} _{(1)}=0\,.
\end{equation}
Therefore, $\tilde{H}^{(n)}$ reduces to 
\begin{equation}
\tilde{H}^{(n)}=c_{\mathcal{J}}\left(\frac{\partial H_{\text{KdV}}^{(n)}}{\partial c_{\mathcal{P}}}+\int d\phi\frac{\partial\mathcal{P}}{\partial c_{\mathcal{P}}}\frac{\delta H_{\text{KdV}}^{(n)}}{\delta\mathcal{P}}\right)+at\dot{H}_{\text{KdV}}^{(n)}\,,\label{eq:Htilden}
\end{equation}
where the last term in \eqref{eq:Htilden} vanishes since $H_{\text{KdV}}^{(n)}$
is conserved. Hence, \eqref{eq:Htilden} can be written in terms of
the total derivative with respect to the central extension $c_{\mathcal{P}}$,
according to
\begin{equation}
\tilde{H}^{(n)}=c_{\mathcal{J}}\frac{dH_{\text{KdV}}^{(n)}}{dc_{\mathcal{P}}}\,.
\end{equation}

\providecommand{\href}[2]{#2}\begingroup\raggedright\endgroup

\end{document}